\documentclass[aps,pra,superscriptaddress,showkeys,floatfix,twocolumn,nobibnotes,nofootinbib]{revtex4}

\usepackage{amsmath,amssymb}    % need for subequations
\usepackage{amsfonts} % fonts
\usepackage{graphicx}   % need for figures
\usepackage{verbatim}   % useful for program listings
\usepackage{xcolor}     % use if color is used in text
\usepackage{subfigure}  % use for side-by-side figures external documents and URLs

\newcommand{\eff}{\text{eff}}
\newcommand{\TE}{\text{TE}}
\newcommand{\TM}{\text{TM}}
\newcommand{\TO}{\text{TO}}

\newcommand{\mum}{\mu \text{m}}

\begin{document}
	
	%\title{Mie scattering by polaritonic cylinders}
	\title{Polaritonic cylinders as multifunctional  metamaterials: Single scattering and effective medium description} 
	\author{Charalampos P. Mavidis}
	\email{mavidis@iesl.forth.gr}
	\affiliation{Department of Materials Science and Technology, University of Crete, Heraklion, Crete, Greece}
	\affiliation{Institute of Electronic Structure and Laser, Foundation for Research and Technology Hellas, N. Plastira 100, 70013 Heraklion, Crete, Greece}
	\author{Anna C. Tasolamprou}
	\affiliation{Institute of Electronic Structure and Laser, Foundation for Research and Technology Hellas, N. Plastira 100, 70013 Heraklion, Crete, Greece}

	\author{Eleftherios N. Economou}
	\affiliation{Institute of Electronic Structure and Laser, Foundation for Research and Technology Hellas, N. Plastira 100, 70013 Heraklion, Crete, Greece}
	\affiliation{Department of Physics, University of Crete, Heraklion, Greece}
	
	\author{Costas M. Soukoulis}
	\affiliation{Institute of Electronic Structure and Laser, Foundation for Research and Technology Hellas, N. Plastira 100, 70013 Heraklion, Crete, Greece}
	\affiliation{Ames Laboratory and Department of Physics and Astronomy, Iowa State University, Ames, Iowa 50011, USA}
	\author{Maria Kafesaki}
	\affiliation{Department of Materials Science and Technology, University of Crete, Heraklion, Crete, Greece}
	\affiliation{Institute of Electronic Structure and Laser, Foundation for Research and Technology Hellas, N. Plastira 100, 70013 Heraklion, Crete, Greece}
	
	% ABSTRACT
	\begin{abstract}
		Polaritonic materials, owing to a strong phonon-polariton resonance in the THz and  far-infrared parts of the electromagnetic spectrum, offer both high-index dielectric and metallic response in this regime. This complex response makes them suitable candidates for the design of metamaterial-related  phenomena and applications.
		Here we show that one type of polaritonic-material-based structures that are particularly suitable for the achievement of a wide range of metamaterial properties are systems of polaritonic rods. To study the interplay between the material and the structural resonances in such systems we employ as model systems rods of LiF and SiC and we calculate first the scattering properties of a  single  rod, identifying and discussing the behavior of the different resonances for different rod diameters.
		To analyze the response of ensembles of polaritonic rods we employ an effective medium approach based on the Coherent Potential Approximation (CPA), which is shown to be superior to the simple Maxwell-Garnett approximation for polaritonic and high-index dielectric metamaterials.
		Calculating and analyzing the CPA effective parameters, we found that our systems  exhibit a large variety of interesting metamaterial properties, including hyperbolic dispersion, epsilon-near-zero and negative refractive index response. This rich response, achievable in almost any system of polaritonic rods, is highly engineerable by properly selecting the radius and the filling ratio of the rods, making polaritonic rod systems an ideal platform for demonstration of multifunctional metamaterials.
	\end{abstract}
	
	% KEYWORDS
	%\date{\today}
	
	\maketitle
	
	% INTRODUCTION
	\section{INTRODUCTION}
	\label{sec:intro}
	
	The emergence of electromagnetic (EM) metamaterials (MM), i.e. engineerable structured materials made of sub-wavelength resonant building blocks (meta-atoms) with novel and unique EM properties and response, made possible the demonstration of novel and unconventional EM wave phenomena, entailing possibilities to advance or even revolutionize a great variety of applications related with EM wave control, from telecommunications, to imaging, sensing etc. 
	%The majority of the metamaterial-originated phenomena are determined mainly from the geometry of the structure building blocks (metalic, dielectric or semiconducting, acting as EM resonators); properly designing those building blocks one can achieve almost on demand EM properties and control.
	Particularly interesting categories of metamaterials that have been designed and demonstrated so far include: (a) Negative effective permeability (mu-negative, MNG) and negative refractive index metamaterials (NIMs, usually achievable by combining negative effective permittivity and permeability)~\cite{Veselago1968SP,Pendry1999IEEE}. NIMs are associated with many counter intuitive phenomena, such as opposite phase and energy velocity, negative refraction etc., and unique potential in imaging and telecommunications applications.
	The first realizations of MNG and NIM structures were obtained employing and properly structuring metals~\cite{Shelby2001Science}, while, later, it was shown that the same response can be achieved also by metamaterials made of high-index dielectrics~\cite{Peng2007PRL, Vynck2009PRL}, where the strong displacement current undertakes the role that conduction current plays in metals. 
	(b) Hyperbolic metamaterials (HMMs)~\cite{Liu2008OE,Poddubny2013NatPhot}, i.e. anisotropic metamaterials showing hypebolic dispersion relation, own to the mixed positive and negative values of their effective permittivity or permeability tensor components. Such metamaterials, which are usually realized by properly alternating metallic and dielectric layers or by employing metallic rod systems, show great potential in imaging applications~\cite{Kim2018SciRep,Habib2019PRB}, as they can offer almost perfect imaging, even with magnification (they can  transform evanescent waves to propagating), and in spontaneous emission enhancement~\cite{Jacob2012APL,Lu2018AdvMat} (as they can offer very high density of EM states).
	(c) Metamaterials with permittivity near zero (ENZ)~\cite{Silveirinha2006PRL,Maas2013NatPhot,Briere2016OL}; such metamaterials, which can be realized by properly engineering electrical permittivity resonances (e.g. by proper structuring), are associated with peculiar phenomena and possibilities, e.g. squeezing of EM waves in very narrow channels, easy wavefront engineering, etc.  Moreover,  the huge wavelength in such metamaterials makes them ideal hosts for demonstration of subwavelength phenomena, as it makes all the wave propagation and scattering features in to them to fall in the extreme sub-wavelength region, almost for any type of embedded scatterers.  
	\begin{figure}[tb]
		\begin{center}
			\includegraphics[width=86mm]{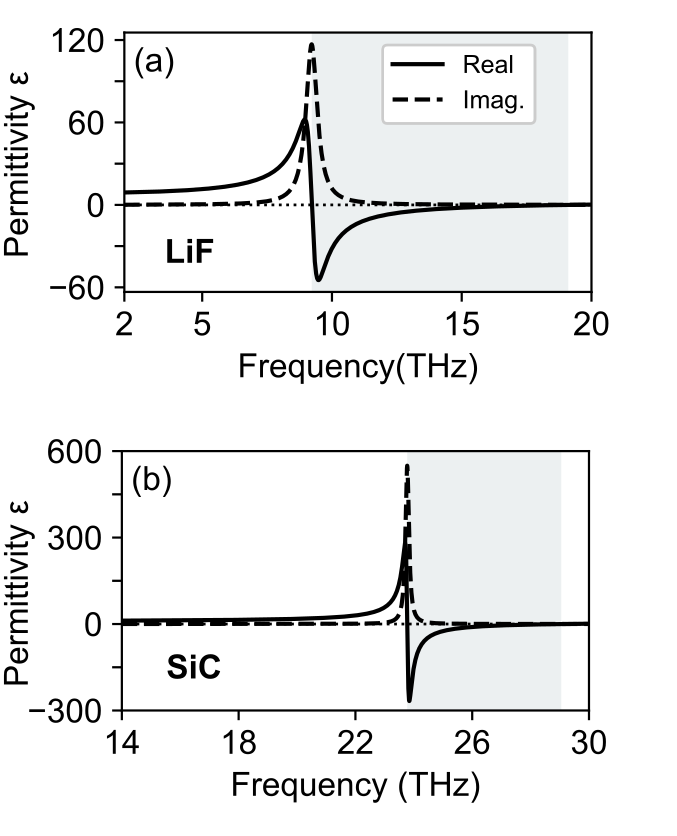}
			\caption{\label{fig:LiFSiC} Real and imaginary parts of the dielectric function for (a) LiF and (b) SiC calculated from Eq. (\ref{eq:Lorentz}) and using the data from Table \ref{tab:materials}. The gray areas indicate the frequency regions where the real part of the corresponding dielectric function is negative, between approximately $\omega_T$ and $\omega_L$.
			}
		\end{center}
	\end{figure}
	As we aim to show in this article, all the above metamaterial categories and their related novel phenomena are achievable with properly  engineered systems of phonon-polariton materials (polaritonic systems)~\cite{Foteinopoulou2019Nanophot}, in particular in  systems made of polaritonic cylinders in a dielectric host.
	%i.e. of cylinders made of materials with phonon-polariton resonances. 
	
	% in In this article we aim to show that periodic systems made of phonon-polaritonic cylinders can be engineered to exhibit all the above mentioned metamaterial properties and capabilities.
	
	% (d) Metasurfaces (MSs), i.e. thin metamaterial layers. The possibility of modulation of the structure-geometry (and thus the resonant response of the meta-atoms) along the metasurface~\cite{Chen2016RepPhys,Tsilipakos2018AOM,Liu2019PRAP} offers to MSs a much greater potential than that of just thin MM layers; indeed anomalous refraction, focusing, perfect transmission or reflection, perfect polarization conversions and many other interesting effects have been demonstrated employing metasurfaces.
	
	% As was mentioned also earlier, the first metamaterials were obtained employing metallic building blocks, while, later, metamaterials made of high index dielectrics~\cite{Jahani2016NatNano,Butakov2016SciRep} showed (and show) in many cases superior response, especially for operation in the optical region where metals are associated with very high ohmic losses. 
	
	Phonon-polariton (polaritonic) materials~\cite{Huang2004APL,Foteinopoulou2011PRB,Foteinopoulou2019Nanophot} is a particularly interesting category of materials, combining both metallic and dielectric response. They are polar crystals (e.g. NaCl) where the EM radiation excites lattice vibrations, resonant in the region from THz to far- and mid-IR. The coupling of the EM radiation with the lattice vibrations in that region results to a resonant permittivity response of  Lorentz-type, i.e.
	\begin{equation}
	\label{eq:Lorentz}
	\varepsilon ( \omega )  = \varepsilon_\infty\frac{\omega^2-\omega_L^2+i\omega^2\Gamma}{\omega^2-\omega_T^2+i\omega^2\Gamma}
	\end{equation}
	where $\varepsilon ( \omega )$ is the relative permittivity,  the resonance frequency $\omega_\TO$ is the transverse optical phonon  frequency, $\Gamma$  is the collision frequency, $\omega_L$ is the longitudinal optical phonon frequency, at which the dielectric function practically vanishes ($\omega_L$ is the analogue of the bulk plasmon frequency of the metallic case) and $\varepsilon_\infty$ stands for the asymptotic value of the relative permittivity at high frequencies (much higher than $\omega_L$ and lower than the frequencies of the inter-band electronic excitations).

	The permittivity for two characteristic polaritonic materials, namely LiF and SiC, is plotted in Fig.~\ref{fig:LiFSiC}. 
	Examining the permittivity forms of Fig.~\ref{fig:LiFSiC}, one can easily realize the great potential of the polaritonic materials in MM-related phenomena and applications.  Polaritonic materials offer regions of  (a) high positive permittivity and thus they can be used for designing and demonstration of any kind of dielectric metamaterials~\cite{Jahani2016NatNano} and metasurfaces; (b) negative permittivity, similar to that of metals in the optical region (with smaller loss-tangent); thus they can provide all the properties and possibilities that metallic metamaterials offer in optics, e.g. plasmonic effects, hyperbolic metamaterial response; (c) permittivity near zero, offering a convenient alternative to complex metamaterial structures that are usually designed to achieve epsilon-near-zero (ENZ) response; besides, they can act as bulk ENZ hosts for demonstration of uncommon scattering and propagation effects~\cite{Liberal2017PHIL}.
	
	An additional merit of the polaritonic materials is that the above mentioned  rich response is exhibited in the THz and far-IR region of the electromagnetic spectrum, a region particularly interesting for sensing, security, biological and medical imaging, and thermal management, and also a region where there is considerable lack of advanced optical components (e.g. the THz gap). Finally, since many of the polaritonic materials are semiconducting, their properties and response can be highly tunable, e.g.  by photoexcitation~\cite{Foteinopoulou2019Nanophot}.
	
	The potential of the polaritonic materials in MMs-related applications makes important the development or adaptation of not only  advanced computational tools suitable for the study of such materials  but also  of simplified models able to explore, identify, explain and even predict the rich variety of phenomena and possibilities allowed by those materials.
	Such a category of simplified models are the well known effective medium models, describing metamaterials as homogeneous (effective) media. The most well established such model is the Maxwell-Garnett (MG)~\cite{MGarnett1904RS} model, suitable for the designing and description of  structures in the quasistatic region.
	The MG model has been extensively applied for either the prediction or the analysis of the metamaterial response of many different structures, especially of structures composed of metallic scatterers of spherical or cylindrical shape, in the low-frequency limit, and specifically when $k_h R\ll 1$, $k_s R\ll 1$, with $k_h$, $k_s$, the wavenumber in the host and scattering material, respectively, and  $R$ the scatterers radius. In the case of systems though made of polaritonic scatterers, as well as in systems of high permittivity dielectric scatterers~\cite{Schuller2009OE,Butakov2016SciRep}, the high permittivity of the scatterers (resulting to small associated wavelength) leads to scattering resonances also in the long-wavelength region (i.e. resonances in the region $k_h R\ll 1$) the influence of which, although crucial for the wave propagation, can not be described by the simple quasistatic MG model. As a result, important features of polaritonic or high-index dielectric systems, such as magnetic response by non-magnetic scatterers, cannot be reproduced. 
	To overcome this problem extended MG models have been developed (valid in the region $k_h R\ll 1$, $k_s R\approx 1$) and have been applied with great success in  systems made of spherical scatterers, either polaritonic or high-index dielectric~\cite{Yannopapas2005JPCM,Yannopapas2007APA}.
	For the case of cylindrical scatterers~\cite{OBrien2002JPCM,Schuller2007PRB,Zhang2015SciRep}, though, the most well-known suitable effective medium description is a description based on field homogenization~\cite{OBrien2002JPCM}, which is not straightforward to apply, while extended Maxwell-Garnett approaches, to our knowledge, have not been developed and applied in detail up now. In this paper, we show that a homogenization approach based on the well known in the Solid State Physics community Coherent Potential Approximation (CPA) method~\cite{Wu2006PRB,PSheng2006book} can be applied with great success in the case of polaritonic rod systems, demonstrating a variety of novel and unconvenional metamaterial phenomena in such systems. 
	\begin{table}[b]
		\caption{\label{tab:materials} Lorentz model material  parameters for LiF and SiC.}
		\begin{ruledtabular}
			\begin{tabular}{lcccc}
				\textbf{Material}   & $\varepsilon_\infty$ & $\omega_T/2\pi$ (THz) & $\omega_L/2\pi$ (THz) & $\Gamma/2\pi$ (THz)\\
				\hline
				LiF \cite{Coronado2012OE} & 2.027 & 9.22 & 19.11 & 0.527\\
				SiC \cite{Hillenbrand2004UM} &  6.7 & 23.79 & 29.05 & 0.143 
			\end{tabular}
		\end{ruledtabular}
	\end{table}
	% END TABLE
	We have to note here that various systems of polaritonic rods in a host  have been already studied, not only theoretically but also experimentally, and interesting phenomena and possibilities have been predicted or demonstrated:
	%have been proven as a model system for  demonstration of many of the  capabilities of polaritonic materials in MMs-related phenomena and applications: 
	It has been shown that by properly designing the radii, heights and distances of the rods, one can achieve both negative permeability and negative refractive index response~\cite{Schuller2007PRB}. Moreover, hyperbolic response in such systems has been already theoretically demonstrated~\cite{Foteinopoulou2011PRB,Coronado2012OE}, toroidal dipolar response~\cite{Tasolamprou2016PRB,PhysRevB.100.085431}, epsilon-near-zero originated waveguiding ~\cite{Masaouti2013OL}, and other interesting and useful effects. 
	Moreover the possibility to relatively easily obtain such systems by, e.g. eutectics self-organization~\cite{Pawlak2010AdvFMat,Coronado2012OE}, laser micromachining~\cite{Ward2007APA}, etc., makes their study even more appealing and indispensable.
	
	The aim of this paper is to analyze in detail the wave propagation in systems of circular polaritonic rods (of infinite height) in a dielectric host and to identify the different interesting propagation regions and their associated characteristics. Of particular interest is the investigation and analysis of the effect of the combination of the material resonances (such as those shown in Fig.~\ref{fig:LiFSiC}) with the structure resonances, dependent on the shape and size of the rods.
	To that extent, the approaches and many of the results of the paper are not  applicable only in the case of phonon-polariton systems but they can be applied in any system made of scatterers from a resonant material (e.g. exciton-polariton systems, macroscopic MMs forming cylindrical scatterers, etc.); moreover, the results can be transferred easily in  the case of high-index dielectric scatterers~\cite{Fu2013NatComm,Tasolamprou201423147,Staude2017NaturePhot,Tasolamprou2017ACSPhot,Mavidis2020PRB}.
	
	To analyze the response of the polaritonic rod systems and to understand the effect of the interplay of material and structure resonances we start from calculation and analysis of the single rod extinction and scattering cross section; then we use the single rod results in the application of the CPA approach~\cite{Wu2006PRB,Kafesaki19977,Kafesaki1998383}, which is employed for the investigation and analysis of the multirod systems. As model systems we employ two different polaritonic rod systems: systems made of LiF rods and systems made of SiC rods (see Fig.~\ref{fig:LiFSiC} for the materials permittivity).
	\begin{figure}[tb]
		\begin{center}
			\includegraphics[width=8cm]{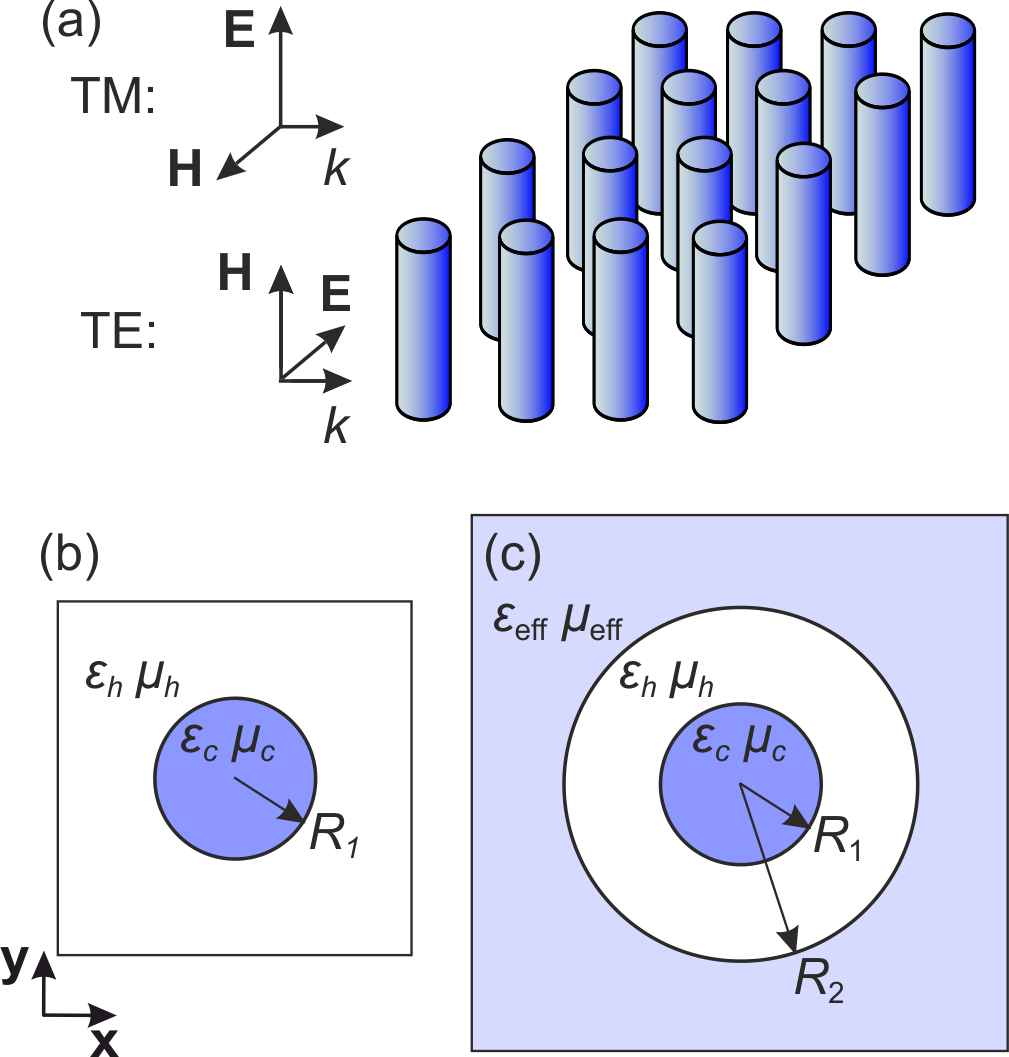}
			\caption{\label{fig:coated} (a) An assembly of cylinders in a host medium (our system of interest) and definitions of the TE and TM polarization and the normal to the cylinders axes plane of incidence. (b) The unit cell of the system of (a), along with its relevant geometry and material parameters, i.e. electrical permittivity, $\varepsilon$, and magnetic permeability $\mu$. The subscripts $h$, $c$ stand for the host and cylinder material respectively. (c) Geometry for the derivation of the effective electric permittivity, $\varepsilon_{\eff}$, and effective magnetic permeability, $\mu_{\eff}$: A single cylinder of radius $R_1=R$  coated  by a coating of thickness $R_2-R_1$ made of the host material of the original system, embedded in the effective medium. $R_2$ is such as $f=R_1^2/R_2^2$, where $f$ is the cylinder filling ratio in the original system.}
		\end{center}
	\end{figure}
	Specifically, the paper is organized as follows: In Sec.~\ref{sec:methods} we introduce the methods used for the calculation of the single rod extinction efficiencies and the relations for the effective medium determination. In Sec.~\ref{sec:results} we present the results of single rod scattering (subsection~\ref{subsc:singlesc}) and of the effective medium (subsection~\ref{subsc:effective}) for our particular systems and we identify the different attainable interesting MM properties and capabilities. Comparison of our results with full-wave simulations demonstrate and verify the validity and merit of our approach in the study of polaritonic and high-index dielectric MMs, validating also further the feasibility  of the interesting attainable effects predicted.

	%Metamaterial reviews: \citep{SoukoulisMeta2011}
	
	%Hyperbolic: \cite{HyperbolicReview2012, PurcellHyperbolic}
	
	%optical dielectric metamaterials: %\cite{MagnLight2012,OptMagn2012,isotropicOptical2013}

	\section{Methods}
	\label{sec:methods}
	
	Although the systems of interest in this work are systems of polaritonic rods in air or in a dielectric host,  the methods  discussed in this section are derived for a general system of (identical) rods in a host, allowing any permittivity and permeability for both the rod and the host material. This is in order to achieve the widest possible applicability regime of the derived formulas, allowing their use for prediction or understanding of the properties of other potentially interesting MM systems or categories.
	
	% SINGLE SCATTERING
	\subsection{Single Scattering }
	\label{sec:singlesc}
	We consider a single infinitely-long cylinder~\cite{Bohren1998Book,Stratton2015Book} with radius $R$, composed of a material with relative electrical permittivity $\varepsilon_c$ and magnetic permeability $\mu_c$ embedded in a host material with material parameters $\varepsilon_h$ and $\mu_h$.  Along the rest of the paper the subscripts $h$ and $c$ in any quantity would refer to host and cylinder respectively. Moreover we consider propagation in a plane perpendicular to the cylinder axis.
	Since the cylinder is infinitely-long and there is no propagation component parallel to its axis, the problem is two dimensional and, due to symmetry, it can be decoupled into two separate polarizations, the Transverse Electric (TE) polarization, with the electric field normal to the cylinder axis, and the Transverse Magnetic (TM) polarization, with the magnetic field normal to the cylinder axis, as seen in Fig.~\ref{fig:coated}(a).  The fields can be expanded on the basis of cylindrical harmonics inside and outside of the cylinder and the expansion coefficients can be found by imposing the appropriate boundary conditions on the cylinder's surface ~\cite{Jackson1999Book}. 
	Specifically, the parallel to the cylinder axis component of the scattered magnetic/electric field is proportional to $\sum_{m=-\infty}^{\infty}a_m^\mathtt{P} \mathbf{N}_{em,k_h}$ where $\mathbf{N}_{em,k_h}=k_h H_m(k_h \rho) \cos(m\varphi) \hat{\mathbf{z}}$ denotes the $m$-th order cylindrical harmonic and the coefficient $a_m^\mathtt{P}$ denotes the Mie scattering coefficient of the $m$-th mode for polarization $\mathtt{P} = \lbrace{\text{TE}, \text{TM} \rbrace}$, which is given~\cite{Bohren1998Book,Stratton2015Book} by
	%\begin{widetext}
	\begin{equation}
	\label{eq:Hcoeff}
	a_m^{\text{TE}} = \frac{ \eta_h J_m(k_cR) J_{m}'(k_hR) - \eta_c J_m(k_hR)J_{m}'(k_cR)}{\eta_c J_m'(k_cR)H_{m}(k_hR)-\eta_h H_m'(k_hR)J_{m}(k_cR)  }
	\end{equation}
	\begin{equation}
	\label{eq:Ecoeff}
	a_{m}^{\text{TM}}  = \frac{\eta_h  J_{m}'(k_cR)J_{m}(k_hR) - \eta_c  J_{m}'(k_hR)J_{m}(k_cR)}{\eta_c  J_m(k_cR) H_{m}'(k_hR) - \eta_h  H_m(k_hR) J_{m}'(k_cR) }
	\end{equation}
	%\end{widetext}
	where $k_{h}=\sqrt{\varepsilon_h \mu_h} \omega/c$  is the wavenumber in the host material,   $k_{c}=\sqrt{\varepsilon_c \mu_c} \omega/c$  is the wavenumber in the cylinder and $\eta_c=\sqrt{\mu_c/\varepsilon_c}$, $\eta_h=\sqrt{\mu_h/\varepsilon_h}$ denote the impendances of the cylinder and the host material respectively.  $J_m$ and $H_m$  stand for the Bessel and Hankel function (respectively) of the first kind and order $m$, and $J'_m$ and $H'_m$ are their derivatives in respect to their argument. 
	
	The dominant modes for each case can be identified from the extinction efficiency, $Q_\text{ext}$, which is defined as the sum of the electromagnetic field energy scattered and absorbed by the cylinder, normalized to the incident energy and the geometric cross section of the cylinder, $2R$. In terms of the scattering coefficients it can be written as\footnote{In literature  ~\cite{Bohren1998Book, Schuller2009OE} there is no minus sign in the extinction efficiency; it is due to  the definition of the coefficients $a_m$ with an extra minus sign (see Appendix~\ref{app:affective-Derivation}).}
	\begin{equation}
	Q_{\text{ext}}^{\mathtt{P}} =  -\frac{2}{|k_h R|} \texttt{Re}\left [ a_0^{\mathtt{P}} + 2\sum_{m=1}^{\infty} a_m^{\mathtt{P}} \right ] \label{eq:Qext} 
	\end{equation} 
	and it can be decomposed into scattering efficiency, $Q_{\text{sc}}$, and absorption efficiency, $Q_{\text{abs}}$, given by
	
	\begin{equation}
	Q_{\text{sc}}^{\mathtt{P}} =  \frac{2}{|k_h R|} \left[ |a_0^{\mathtt{P}}|^2 + 2\sum_{m=1}^{\infty} |a_m^{\mathtt{P}}|^2 \right] \label{eq:Qsc} 
	\end{equation} 
	\begin{equation}
	Q_{\text{abs}}^{\mathtt{P}}=Q_{\text{ext}}^{\mathtt{P}}-Q_{\text{sc}}^{\mathtt{P}} 
	\end{equation}
	The resonances in the extinction spectra of the cylinders (also known as Mie resonances~\cite{Mie1908AnnPhys}) can be classified by their polarization $\mathtt{P}$ and an integer $m$ associated with the corresponding cylindrical harmonic. The resonance frequencies or eigenfrequencies of the system can be calculated by setting the denominators of the scattering coefficients for each polarization, Eq.~(\ref{eq:Hcoeff}) for TE and Eq.~(\ref{eq:Ecoeff}) for TM, equal to zero: 
	\begin{equation}
	\label{eq:Hresonance}
	\frac{1}{\eta_c} \frac{J_m (k_c R)}{J'_m (k_c R)} = \frac{1}{\eta_h} \frac{H_m (k_h R)}{H'_m (k_h R)}
	\end{equation}
	\begin{equation}
	\label{eq:Eresonance}
	\eta_c \frac{J_m (k_c R)}{J'_m (k_c R)} = \eta_h \frac{H_m (k_h R)}{H'_m (k_h R)}
	\end{equation}
	
\begin{figure*}[tb]
		\begin{center}
			
			\includegraphics[width=17cm]{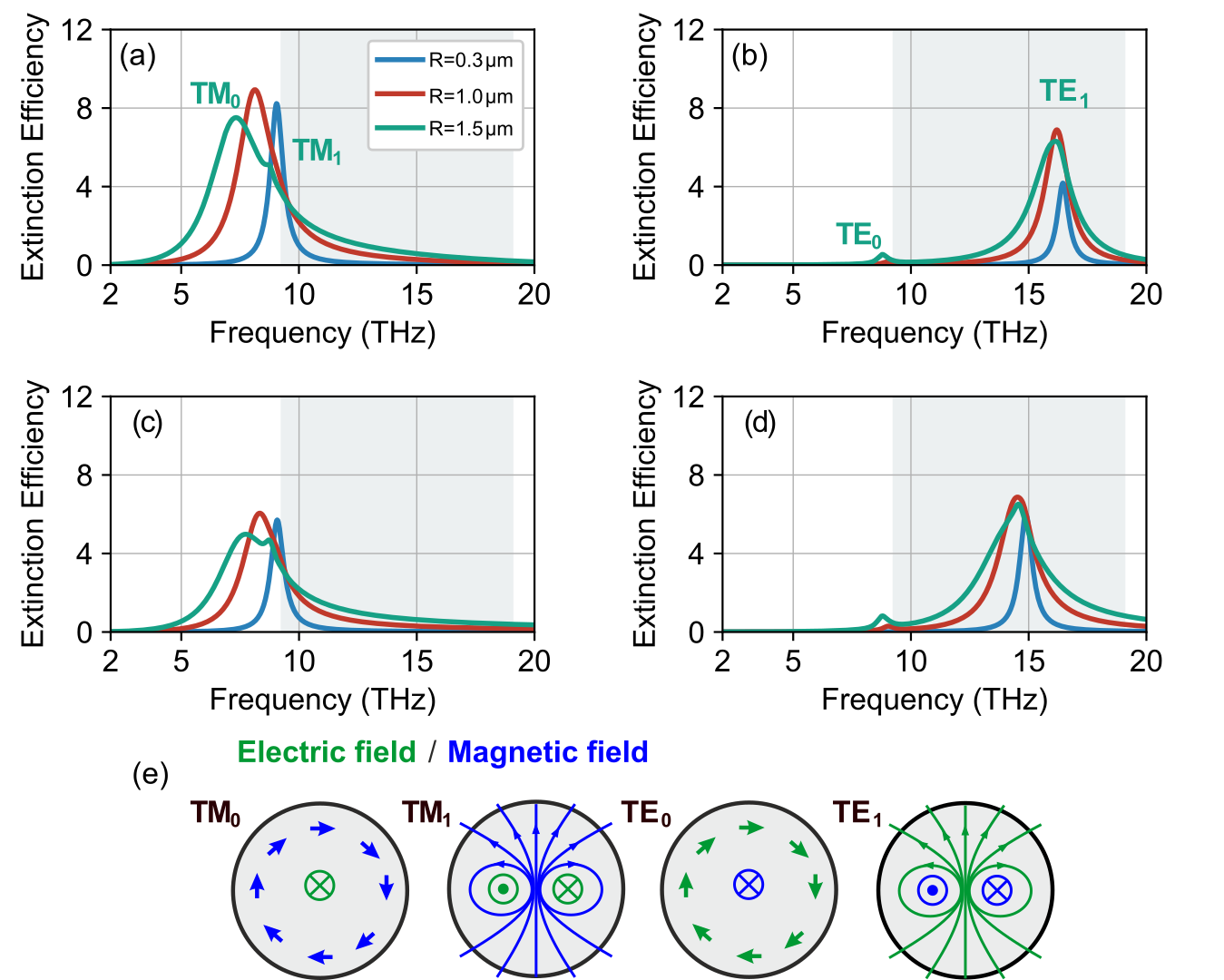}
			\caption{\label{fig:LiFext} Extinction efficiency of a LiF cylinder in air [(a),(b)] and in a dielectric with $\varepsilon_h=2$ [(c),(d) for TM (left column) and TE (right column) polarizations. The legend shows the cylinder radius. The shaded areas correspond to the frequency region where the dielectric function of LiF is negative. (e) Electric (green color) and magnetic (blue color) field distributions  for the TM$_0$, TM$_1$, TE$_0$ and TE$_1$ modes.
			}
		\end{center}
	\end{figure*}

	%{\color{purple}The solutions of Eq.~(\ref{eq:Hresonance}) and  Eq.~(\ref{eq:Eresonance}) are each identified by two integers  $m$ and $n$ which correspond the order of the  the cylindrical harmonics, $m$,  and  the number of roots of the transcendental Bessel functions , $n$. The identifier $m$ is associated with the azimuthal spatial profile of the modes and the  identifier $n$ is associated with the radial spatial profile of the modes. The radial variation of the modes is met in higher order natural modes; in the present study, where only the fundamental excitations are involved we omit the identifier $n$ ($n=0$) and we symbolically write each mode as $\mathtt{P}_m$.}
	In the limit $k_h R \ll 1$ (where a system of cylinders behaves as a metamaterial) it is sufficient to consider only the first  two fundamental modes, i.e.  $m=0$ and $m=1$, since the contribution of higher order modes is insignificant. In the discussion below, these modes are identified as TE$_0$, TE$_1$, TM$_0$ and TM$_1$ (for the fields distribution of those modes see Fig.~\ref{fig:LiFext}). Using  recurrence and other relations of Bessel functions \cite{Stegun} (e.g. $J'_0(x) = -J_1(x)$ and $H'_0(x) = - H_1(x)$) we see that for $\mu_h=\mu_c$ the eigenfrequency relations of the TE$_0$ and the TM$_1$ modes are identical; therefore, TE$_0$ and the TM$_1$ modes are degenerate.
	
	\begin{figure}[th]
		
		\begin{center}
			\includegraphics[width=86mm]{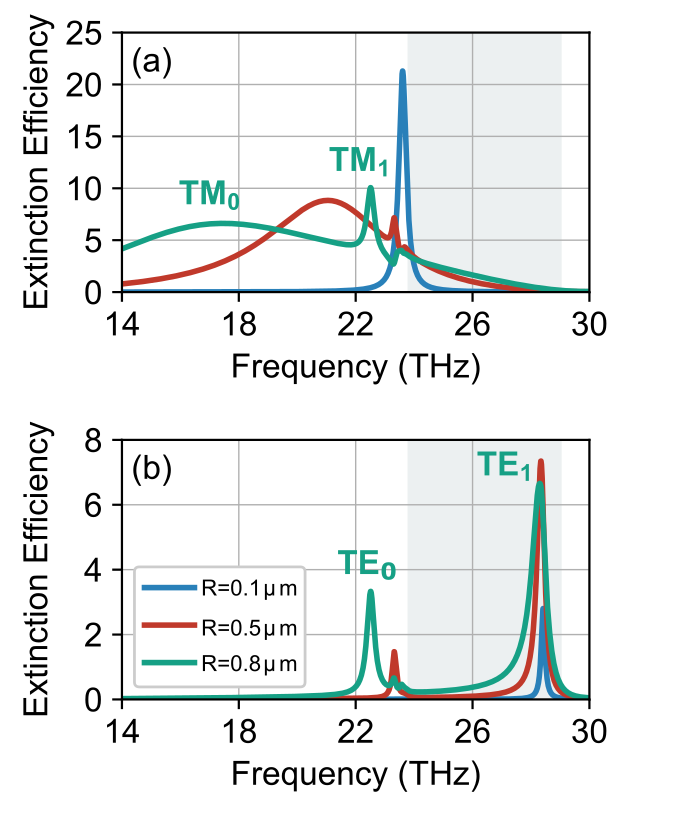}

			\caption{\label{fig:SiCext} Extinction efficiency of a SiC cylinder (of radius 0.1, 0.5 and 0.8 $\mu$m) in air for (a) TM and (b) TE polarizations. The shaded areas correspond to the frequency region where the dielectric function of SiC is negative.
			}
		\end{center}
	\end{figure}
	To explore the eigenfrequency relations of the above modes in the limits of small size parameters $k_h R$ and $k_c R$ we use the limiting expressions of Bessel functions~\cite{Stegun} listed in Appendix~\ref{app:affective-Derivation}. For the TM$_1$ mode, in the limit of $k_h R \ll 1$ we have 
	\begin{equation}
	\label{eq:tm1-limit1}
	\eta_c \frac{J_1 (k_c R)}{J'_1 (k_c R)} = -\eta_h k_h R = -\mu_h \frac{\omega}{c}R 
	\end{equation}
	In the quasistatic limit of both $k_h R \ll 1$ and  $k_c R \ll 1$ the TM$_1$ resonance condition becomes
	\begin{equation}
	\label{eq:tm1-limit2}
	\mu_c=-\mu_h
	%\xrightarrow[\mu_c=\mu_h=0]\left(\ k_h R\right)^2 = -2.
	\end{equation}
	Thus, the TM$_1$ mode does not present resonances in the quasistatic limit, except in the case of a magnetic host or cylinder. 
	For the TE$_0$ in the limit of $k_h R \ll 1$ we have 
	\begin{equation}
	\label{eq:te0-limit1}
	\frac{1}{\eta_c}\frac{J_0(k_c R)}{J'_0(k_c R)} = \varepsilon_h\left[ \ln\left(\frac{k_h R}{2}\right)+\gamma -i\frac{\pi}{2}\right]\frac{\omega}{c}R 
	% \frac{1}{\eta_c} \frac{J_0 (k_c R)}{J'_0 (k_c R)} = \varepsilon_h \left(\gamma-1-i\frac{\pi}{2}\right)\frac{\omega}{c}R
	\end{equation}
	where $\gamma$ is Euler's constant.
	In the quasistatic limit of both $k_hR \ll 1$ and $k_c R \ll 1$, the resonance condition has solutions only if 
	$1/\mu_c \rightarrow 0$
	in this limit.
	
	In an analogous way one can obtain limiting expressions also  for the TM$_0$ mode; in the limit of $k_h R \ll 1$ 
	\begin{equation}
	\label{eq:tm0-limit}
	\eta_c \frac{J_0(k_c R)}{J'_0(k_c R)} = \mu_h\left[ \ln\left(\frac{k_h R}{2}\right)+\gamma -i\frac{\pi}{2}\right]\frac{\omega}{c}R 
	% \eta_c \frac{J_0 (k_c R)}{J'_0 (k_c R)} = \mu_h\left(\gamma-1-i\frac{\pi}{2}\right)\frac{\omega}{c}R
	\end{equation}
	In the quasistatic limit, of both $k_h R \ll 1$ and  $k_c R \ll 1$, we can have resonance in the case that 
	$1/\epsilon_c \rightarrow 0$, a condition that can be fulfilled in the case of a polritonic cylinder, with the resonance frequency to coincide with the phonon polariton resonance frequency of the cylinder material. 
	
	Finally, for the TE$_1$ mode in the limit of $k_h R \ll 1$  we obtain
	\begin{equation}
	\label{eq:te1-limit}
	\frac{1}{\eta_c} \frac{J_1 (k_c R)}{J'_1 (k_c R)} = -\frac{1}{\eta_h} k_h R=-\varepsilon_h \frac{\omega}{c}R 
	\end{equation}
	In the limit $k_h R \ll 1$ and $k_c R \ll 1$ the resonance condition becomes 
	\begin{equation}
	\label{eq:spplike}
	\varepsilon_c = -\varepsilon_h.    
	\end{equation}
	Equation~(\ref{eq:spplike}), which can be fulfilled in the case of a polaritonic cylinder (owing to its metal-like behavior in frequencies  above the phonon-polariton resonance frequency), is identical to the resonance condition of a Surface Plasmon Polariton (SPP) mode in a dielectric-metal planar interface~\cite{maier2007plasmonics}.

	\subsection{Effective medium}
	\label{sec:effmedium}
	
	We now calculate the components of the effective medium permittivity and permeability tensors for a uniaxial anisotropic system of infinitely long parallel circular cylinders employing a Coherent Potential Approximation (CPA) based approach as developed by Wu~\textit{et. al.}~\cite{Wu2006PRB}. As was already mentioned, unlike the quasistatic Maxwell-Garnett approximation~\cite{MGarnett1904RS}, which is valid only when all $k_h R$, $k_c R$ and $k_{\eff} R$ are much less than unity, the CPA approach allows application in higher frequency regions, where particle resonances occur (and thus interesting metamaterial effects), allowing treatment of metamaterials made of high-index dielectric or polaritonic scatterers.
	A considerable advantage of CPA over other suitable effective medium approaches  (like the field-averaging method~\cite{OBrien2002JPCM}) is that
	the effective parameters are given in a closed form as we will see below. Moreover, the effective parameters do not depend on the specific lattice-type of the system to be described, as it would be in the case of  extended Maxwell-Garnett approaches~\cite{Ruppin2000OptCom}  applied in two-dimensional systems~\cite{DellAnna2016PRA}.
	
	Regarding our systems, as an implication of symmetry, for a proper choice of axes, that is the cylinders are oriented along the $z$ axis, the effective electric permittivity and magnetic permeability must be uniaxial, i.e. diagonal tensors with only two free parameters. 
	In dyadic form they can be written as $\varepsilon_{\eff} =  \varepsilon^{\perp}_{\eff} (\hat{x}\hat{x} + \hat{y}\hat{y}) + \varepsilon^{\parallel}_{\eff} \hat{z}\hat{z}$ and $\mu_{\eff} = \mu^{\perp}_{\eff} (\hat{x}\hat{x} + \hat{y}\hat{y}) + \mu^{\parallel}_{\eff} \hat{z}\hat{z}$ respectively, where $\hat{x}$, $\hat{y}$ and $\hat{z}$ are the unit vectors along the axes, and the symbols $\parallel$ and $\perp$ denote that the corresponding field (electric for  $\varepsilon$ and magnetic for $\mu$) is parallel and perpendicular (respectively) to the cylinders axis.
	
	In order to derive semi-analytical expressions for the components of the effective parameters in the framework of CPA we should require vanishing of the scattering  between effective and actual medium. To apply this requirement we consider the scattering configuration shown in Fig.~\ref{fig:coated}(c), where the actual medium is represented by a coated cylindrical inclusion  (of infinite height) with core of radius $R_1=R$ and material the same as the original cylinders, and coating of thickness $R_2-R_1$ made of the host material of the original system; the coated inclusion is embedded in the homogeneous effective medium with electric permittivity $\varepsilon_{\eff}$ and magnetic permeability $\mu_{\eff}$. The radius $R_2$ of the coated inclusion is defined by the filling ratio, $f$, of the cylinders in the original system, as $f = R_1^2/R_2^2$.  
	In order for the effective medium of Fig.~\ref{fig:coated}(c) to be the one accurately describing our inhomogeneous system CPA requires the scattering cross section from the embedded into the effective medium coated cylinder to be identically zero.
	Hence, all the scattering coefficients of the coated inclusion must be set equal to zero, which, after some algebraic manipulations (see Appendix~\ref{app:affective-Derivation}), leads to the following condition:
	\begin{equation} 
	a_m^{\mathtt{P}}(R_1;c, h) =  a_m^{\mathtt{P}}(R_2;\eff,h) \label{eq:CPA_def}
	\end{equation}
	In Eq.~(\ref{eq:CPA_def}) $a_m^{\mathtt{P}}(R;A,B)$ stands for the $m$-th order scattering coefficient for a cylinder with radius $R$ made of a material $A$ immersed in a host made of material $B$ for polarization $\mathtt{P}$. Equation~(\ref{eq:CPA_def}) defines an infinite system of nonlinear equations which has to be solved  self-consistently for the components $\varepsilon_{\eff}$ and $\mu_{\eff}$ to be obtained.
	However, in the region of $k_h R<1$, which is the region of interest for  metamaterials, we can limit ourselves only to the first two modes per polarization, $m=0$ and $m=1$; then, assuming also that $k_{\eff}R_2\ll 1$ ($k_{\eff}$ is the wavevector norm in the effective medium), we can derive semi-analytical relations for all the components of the effective material tensors (see Appendix~\ref{app:affective-Derivation})
	%~\ref{app:affective-Derivation}), 
	which read as
	%along with the condition where $k_{eff}$ stands for the wavevector norm in the effective medium. 
	\begin{align}
	\varepsilon_{\eff}^{\parallel} &= -\frac{2\varepsilon_h}{k_hR_2 }\left[\frac{J_0'(k_hR_2)+H_0'(k_hR_2)a_0^{\TM}}{J_0(k_hR_2)+H_0(k_hR_2)a_0^{\TM}}\right] \label{eq:cpa1}\\
	\mu_{\eff}^{\perp} &=   \frac{\mu_h}{k_hR_2} \left[ \frac{J_1(k_hR_2) + H_1(k_hR_2)a_1^{\TM}}{J_1'(k_hR_2)+H_1'(k_hR_2)a_1^{\TM}} \right]\label{eq:cpa2} \\
	\mu_{\eff}^{\parallel} &= -\frac{2\mu_h}{k_hR_2} \left[ \frac{J_0'(k_hR_2) + H_0'(k_hR_2)a_0^{\TE}}{J_0(k_hR_2)+H_0(k_hR_2)a_0^{\TE}} \right] \label{eq:cpa3}
	\\
	\varepsilon_{\eff}^{\perp} &=  \frac{\varepsilon_h}{k_hR_2} \left[ \frac{J_1(k_hR_2) + H_1(k_hR_2)a_1^{\TE}}{J'_1(k_hR_2)+H'_1(k_hR_2)a_1^{\TE}} \right] \label{eq:cpa4}
	\end{align}
	where $a_m^{\mathtt{P}} =  a_m^{\mathtt{P}}(R_1;c, h)$. As can be seen in Eqs.~(\ref{eq:cpa1})-(\ref{eq:cpa4}), each one of the effective parameters is related with a particular mode in the single scattering cross section. This connection, justifying the characterization of the modes as electric and magnetic, can be understood also physically by observing the field distribution corresponding to those modes - see Fig.~\ref{fig:LiFresonances} and Section~\ref{subsc:singlesc}.
	
	Equations ~(\ref{eq:cpa1})-(\ref{eq:cpa4}) under certain conditions can lead to resonances in the effective parameters, associated with interesting propagating and scattering effects for the composite structure as we will discuss in the next section. For $k_hR_2\ll 1$ (thus also $k_hR_1\ll 1$) the resonance conditions/frequencies (obtained by setting the denominators equal to zero and employing limiting expressions for the Bessel functions - see Appendix~\ref{app:affective-Derivation}) are approximated as follows.
	\\
	For $\varepsilon_{\eff}^\parallel$ (related to TM$_0$ mode):
	\begin{equation}
	\label{epsparreslim}
	\eta_c\frac{J_0(k_c R_1)}{J_0^{\prime}(k_c R_1)}=\mu_h \ln(R_2/R_1)\frac{\omega}{c}R_1  = -\frac{1}{2}\mu_h \ln(f)\frac{\omega}{c}R_1
	\end{equation}
	For $\mu_{\eff}^{\perp}$ (related to TM$_1$ mode):
	\begin{equation}
	\label{muperpreslim}
	\eta_c\frac{J_1(k_c R_1)}{J_1^{\prime}(k_c R_1)}=\frac{f+1}{f-1}\mu_h \frac{\omega}{c}R_1
	\end{equation}
	For $\mu_{\eff}^\parallel$ (related to TE$_0$ mode):
	\begin{equation}
	\label{muparreslim}
	\frac{1}{\eta_c}\frac{J_0(k_c R_1)}{J_0^{\prime}(k_c R_1)}=\varepsilon_h \ln(R_2/R_1)\frac{\omega}{c}R_1 = -\frac{1}{2}\varepsilon_h \ln(f)\frac{\omega}{c}R_1
	\end{equation}
	For $\varepsilon_{\eff}^{\perp}$ (related to TE$_1$ mode):
	\begin{equation}
	\label{epsperpreslim}
	\frac{1}{\eta_c}\frac{J_1(k_c R_1)}{J_1^{\prime}(k_c R_1)}=\frac{f+1}{f-1}\varepsilon_h \frac{\omega}{c}R_1
	\end{equation}
	One can see that the above relations (\ref{epsparreslim})-(\ref{epsperpreslim}) are very similar with the corresponding conditions for single scattering resonances discussed in the previous subsection. In particular, Eqs. (\ref{muperpreslim}) and (\ref{epsperpreslim}) for low cylinder filling ratio $f$ lead to resonance frequencies very close to those of the corresponding TM$_1$ and TE$_1$, respectively, single cylinder resonances - see Eqs.~(\ref{eq:tm1-limit1}) and (\ref{eq:te1-limit}) respectively.  
	\begin{comment}
	\begin{align}
	\label{eq:kR2-limit}
	\varepsilon_\eff^\parallel &=& \varepsilon_h\left[\frac{(k_hR_2)^2-\left(\frac{4i}{\pi}-(k_hR_2)^2+2i\alpha (k_hR_2)^2 \right) a_0^{\TM}}{(k_hR_2)^2+\left[(\frac{i}{\pi}-2i\alpha)(k_hR_2)^2+1 \right]a_0^{\TM}}\right] \\
	\mu_{\eff}^{\perp} &=& \mu_h\left[ \frac{(k_hR_2)^2+ \left((k_hR_2)^2-2i\alpha (k_hR_2)^2 -\frac{4i}{\pi}\right) a_1^{\TM}}{(k_hR_2)^2+\left((k_hR_2)^2-2i\alpha (k_hR_2)^2 + \frac{4i}{\pi}\right) a_1^{\TM}} \right] \\
	\varepsilon_{\eff}^{\perp} &=& \varepsilon_h \left[ \frac{(k_hR_2)^2+ \left((k_hR_2)^2-2i\alpha (k_hR_2)^2 -\frac{4i}{\pi}\right) a_1^{\TE}}{(k_hR_2)^2+\left((k_hR_2)^2-2i\alpha (k_hR_2)^2+ \frac{4i}{\pi}\right) a_1^{\TE}} \right] \\
	\mu_{\eff}^{\parallel} &=& \mu_h \left[\frac{(k_hR_2)^2-\left(\frac{4i}{\pi}-(k_hR_2)^2+2i\alpha (k_hR_2)^2 \right) a_0^{\TE}}{(k_hR_2)^2+\left[(\frac{i}{\pi}-2i\alpha)(k_hR_2)^2+1 \right]a_0^{\TE}}\right] 
	\end{align}
	\end{comment}
	
	Finally, in the quasistatic limit (i.e. for a $k_h R_2 \ll 1$, $k_h R_1 \ll 1$ and $k_c R_1 \ll 1$), Equations ~(\ref{eq:cpa1})-(\ref{eq:cpa4}) reduce to the well-known MG formulae:
	\begin{align}
	\label{eq:MGappr}
	\varepsilon^{\parallel}_{\eff} &= f\varepsilon_{c} + (1-f)\varepsilon_{h} \\ 
	\mu^{\parallel}_{\eff} &= f\mu_{c} + (1-f)\mu_{h} \\ 
	\varepsilon^{\perp}_{\eff} &= \varepsilon_{h}\frac{(1+f)\varepsilon_{c}+(1-f)\varepsilon_{h}}{(1-f)\varepsilon_{c}+(1+f)\varepsilon_{h}} \\
	\mu^{\perp}_{\eff} &= \mu_{h}\frac{(1+f)\mu_{c}+(1-f)\mu_{h}}{(1-f)\mu_{c}+(1+f)\mu_{h}}
	\end{align}
	
	%{\bf Mpamph nomizw oti axizei na breis th limiting ekfrash twn parapanw gia $k_hR2 \longleftarrow 0$ giati h synthikh plhroitai sta metamaterials kai eimai sigourh oti den tha sou dinei poly diaforetiko apotelesma apo thn rigorous synthikh. Epishs na breis tis limiting ekfrases tv=wn a0,a1 gia $k_hR_1 \longleftarrow 0$.  }

	% RESULTS AND DISCUSSION
	\section{Results and Discussion}
	\label{sec:results}
	\subsection{Single scattering}
	\label{subsc:singlesc}
	\begin{figure*}
		\begin{center}
			\includegraphics[width=171mm]{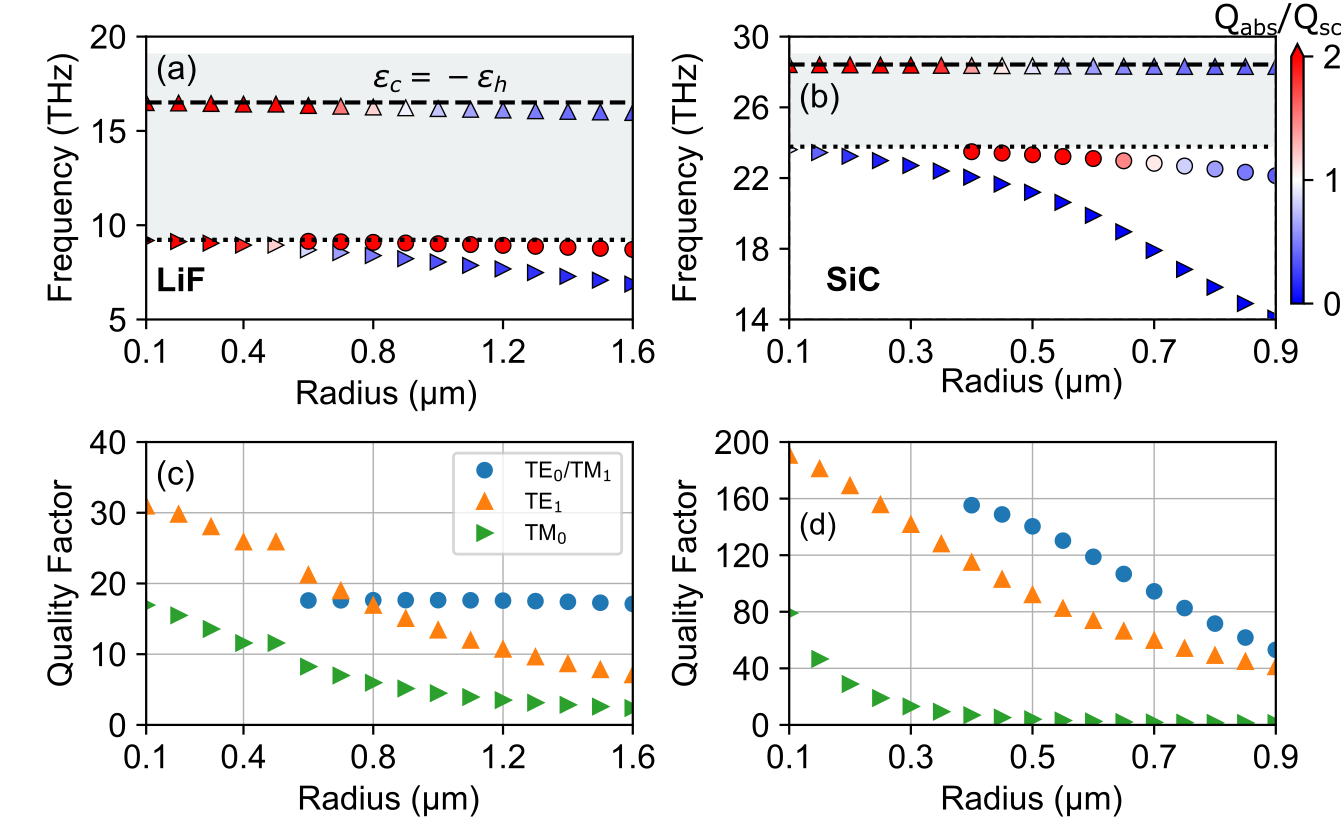}
			
			\caption{\label{fig:LiFresonances} Top panels: Resonance frequencies and absorption over scattering efficiency $Q_\text{abs}/Q_\text{sc}$ (color) at the resonance frequency of  the TM$_0$ (rotated triangles), TE$_0$/TM$_1$ (circles) and the TE$_1$ (upright triangles) modes as function of the radius for a (a) LiF   and a (b) SiC cylinder in air ($\varepsilon_h=1$).  The dashed line shows the quasi-static resonance condition for the TE$_1$ $\varepsilon_c(\omega) = -\varepsilon_h$, and the shaded area corresponds to the frequency region where the dielectric functions of LiF and SiC are negative.  Bottom panels: Quality factor $-\texttt{Re}(\omega_{\text{res}})/(2\texttt{Im}(\omega_{\text{res}}))$ for the modes of a (c) LiF and  (d) SiC cylinder in air.
			}
		\end{center}
	\end{figure*} 
	We begin our analysis by calculating the extinction efficiency of a LiF cylinder in air ($\varepsilon_h=1$, $\mu_h=1$) and in a host with $\varepsilon_h=2$, $\mu_h=1$, and for a SiC cylinder in air, for both TM and TE  polarizations and various radii.
	The dielectric functions of both LiF and SiC, which are shown in Fig. 1, are calculated using Eq.~(\ref{eq:Lorentz}) with parameters  tabulated in Table~\ref{tab:materials}.
	The extinction efficiency results for the LiF and the SiC cylinders are shown in Figs.~\ref{fig:LiFext} and~\ref{fig:SiCext} respectively.
	It is apparent that for each polarization there are two dominant resonances in the low-frequency extinction spectra which originate from the $m=0$ and $m=1$ modes. Using the notation defined in Sec.~\ref{sec:singlesc} we have the TE$_0$, TE$_1$, TM$_0$ and TM$_1$ modes, where the TE$_0$ and TM$_1$ modes resonate at the same frequency, as was also discussed in Section ~\ref{sec:singlesc}. Illustrations of the fields for each of these four modes are shown in Fig.~\ref{fig:LiFext}(c). From the field illustrations one can characterize the modes as electric in nature (i.e. associated with strong induced electric field in the direction of the incoming field),  as TM$_0$ and TE$_1$, and magnetic in nature (i.e. with strong induced magnetic field in the direction of the incoming magnetic field), as the TE$_0$ and TM$_1$.
	
	As can be observed in both Figs.~\ref{fig:LiFext} and~\ref{fig:SiCext}, only the TE$_1$ mode falls in the negative permittivity region of the polaritonic materials (shaded region in the plots) and  is similar in nature to the Localized Surface Plasmon Resonance (LSPR)~\cite{Pfeiffer1974PRB} sustained by metallic particles in the visible part of the spectrum. 
	For very small radii the mode frequency approaches the quasi-static limit ($k_hR\ll 1$ and $k_cR\ll 1$) value, where $\varepsilon_h = -\varepsilon_c(\omega_{\text{res}})$. 
	This relation suggests that the resonance frequency of the TE$_1$ mode is affected greatly by the environment of the cylindrical particle.
	To the contrary, there is no significant dependence of the resonance frequencies of the TE$_0$, TM$_1$ and TM$_0$ modes on the host parameters (only the values of $Q_{\text{ext}}$ change). This result can be partially explained by the fact that we are in the high-index dielectric regime for the cylinder and that the electric fields for these modes are concentrated in or at the surface of the cylinder as related field simulations show. On the other hand this is not true for the TE$_1$ mode, where the electric field is dipole-like and highly extends into the dielectric.
	\begin{figure*}[tb]
		\centering
		\includegraphics[width=171mm]{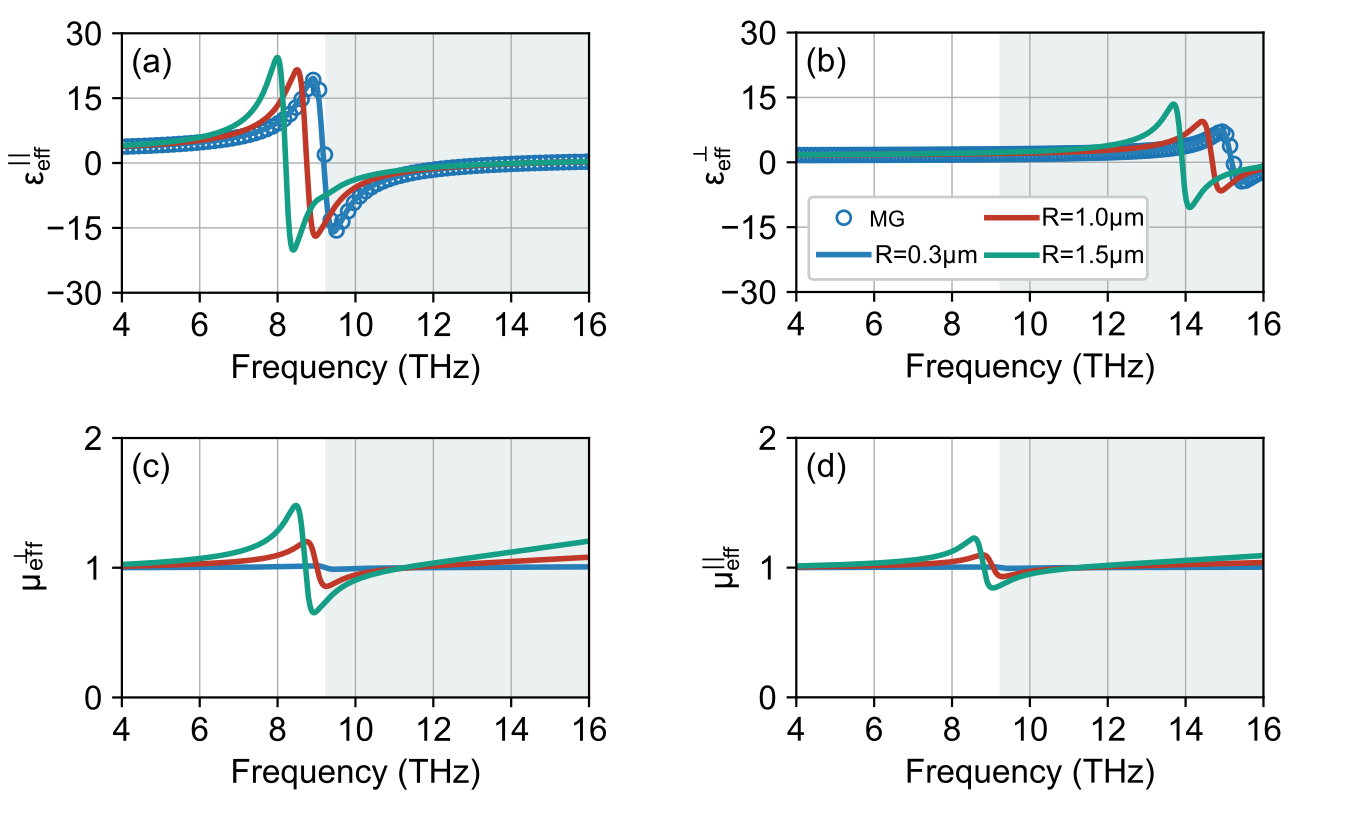}
		
		\caption{\label{fig:LiFeffepsilon} Parallel ((a),(d)) and perpendicular ((b), (c)) components of effective permittivity (upper row) and effective permeability (lower row) for LiF cylinders with filling ratio 30\% in air for different radii using the CPA (lines) and the Maxwell-Garnett  approximation (circles) (only for upper panels). The shaded areas correspond to the frequency region where the dielectric function of the LiF is negative.  }
	\end{figure*}
	
	We turn now our focus on the dependence of resonances on the radius of the polaritonic cylinder. In Fig.~\ref{fig:LiFresonances} we plot the resonance frequencies for a LiF (panel (a)) and a SiC (panel (b)) cylinder in air as a function of cylinder radius, indicating also the ratio of absorption over scattering $Q_\text{abs}/Q_\text{sc}$ of each mode (the resonance frequencies were obtained by solving Eqs.~(\ref{eq:Hresonance}) and~(\ref{eq:Eresonance}) numerically for $m=0$  and $m=1$).
	For small radii only the modes of electric nature appear; i.e. TM$_0$ and  TE$_1$. 
	%For very small radii the mode frequency approaches the quasi-static limit, where $\varepsilon_h = -\varepsilon_c(\omega_{\text{res}})$. 
	For TM polarization, where the incident electric field (parallel to the cylinder) does not experience any "boundaries", the only factor affecting the induced polarization is the polarizabiity of the bulk material; the resulting mode is the spherically symmetric  TM$_0$ mode, with resonance for small radius values almost at the bulk material resonance frequency, $\omega_T$. The resonance frequency moves to lower values as the radius of the cylinder increases and the wavelength inside the cylinder becomes comparable to the radius. This departure of TM$_0$ resonance frequency from $\omega_T$ is faster for the SiC cylinder due to the higher permittivity values and the associated smaller wavelength inside the cylinder.  
	In both TE$_1$ and TM$_0$ modes absorption dominates extinction for small radii, as can be concluded  from Fig.~\ref{fig:LiFresonances}(a), but for larger radii scattering takes over, as happens also in the case of a metallic cylinder. In the SiC case (Fig.~\ref{fig:LiFresonances}(b)) though the dominance of the scattering over absorption for the TM$_0$ mode occurs in very small radius values (even smaller than 0.1$\mum$, which is the threshold value of Fig.~\ref{fig:LiFresonances}(b)), and the absorption cross-section for $R$ larger than 0.2$\mum$ is practically negligible. This can be explained by the quicker departure of the TM$_0$ mode from $\omega_T$ resonance where the losses of SiC are quite high, combined with the much higher quality factor of SiC compared to LiF (note that for SiC $\Gamma/\omega_T \approx 0.006$ while for LiF $\Gamma/\omega_T \approx 0.057$).
	
	The TE$_0$ and TM$_1$ modes, appearing for radius values larger than $0.5\mum$  for LiF and $0.4\mum$ for SiC, appear  also just below  optical phonon frequency $\omega_T$. 
	Their resonance frequency  changes only slightly with the increase of the radius. 
	Moreover, absorption dominates over scattering for small radii and as the radius increases scattering starts to take over. For the case of LiF this happens for radii much larger than those studied here.
	This is probably not-surprising taking into account the weak  extinction cross-section of the TE$_0$ and TM$_1$ modes and the fact that their resonance frequency is (and remains) relatively close to the resonance frequency $\omega_T$ where the material losses are quite high. 
	Indicative plots of LiF and SiC absorption and scattering efficiencies for different cylinder radii are presented in Appendix~\ref{app:SiCscabs}.  
	
	Calculating the quality factor, $\text{Q}$, of the different modes dominating the long wavelength extinction response of LiF and SiC cylinders, with   $\text{Q}=-\texttt{Re}[\omega_{\text{res}}]/(2\texttt{Im}[\omega_{\text{res}}])$, we obtain the result shown in Figs.~\ref{fig:LiFresonances}(c) for LiF and  ~\ref{fig:LiFresonances}(d) for SiC. 
	We observe that for small cylinder radii the quality factor of the TE$_1$ mode, which is sensitive to the environment and thus suitable for sensing applications, gets values higher than 20 for LiF and higher than 100 for SiC cylinders. Such values are higher than the corresponding ones of plasmonic antennas (of the same size parameter, $k_hR$) in the visible~\cite{Wang2006PRL,Zheng2013IEEE,Hrton2020PRAP}, indicating the suitability of polaritonic rods in sensing applications in the THz and IR part of the EM spectrum. 
	Regarding the "magnetic" modes TE$_0$/TM$_1$, for LiF their quality factor changes very slowly with increasing radius and retains values close to 18 (17.59 for $R=0.5\mum$ to 17.12 for $R=1.6\mum$).  For SiC their quality factor decrease with increasing radius occurs much more quickly due to the much lower $\Gamma/\omega_T$ and the quicker departure of the resonance frequency from the highly lossy region around $\omega_T$.

	\subsection{Effective Medium}
	\label{subsc:effective}
	
	\begin{figure}[tb]
		\centering
		\includegraphics[width=86mm]{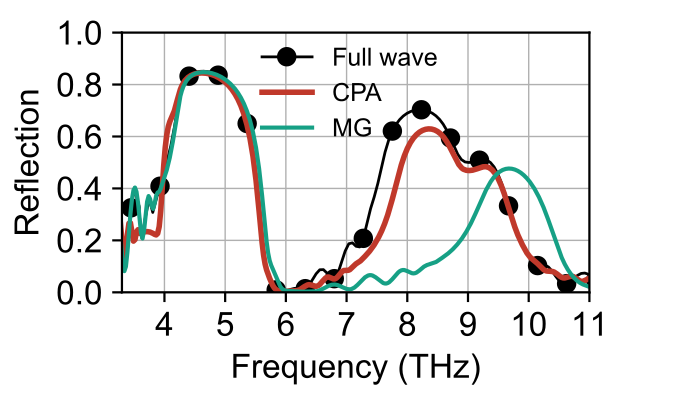}
		
		\caption{\label{fig:comparison} Normal incidence,  TM reflection spectra for a slab of LiF cylinders with radius $R=1.3\mu m$ and filling ratio $f=6.95\%$ in KCl host in a square arrangement. The reflection is calculated by the commercial
			finite element method electromagnetic solver Comsol Multiphysics, considering a computational system  of 7 unit cell thickness (along propagation direction). The full wave reflection results (black line and dots) are compared with results for a homogeneous effective medium of the same thickness as the actual system and effective parameters obtained through CPA 
			(red line) and Maxwell-Garnett approximation (green line). KCl was modelled using Eq.~(\ref{eq:Lorentz}) with parameters  $\varepsilon_\infty=2.045$, $\omega_{T}/2\pi=4.21$THz, $\omega_L/2\pi=6.196$THz and $\Gamma/2\pi = 0.156$THz~\cite{Coronado2012OE}. }
	\end{figure}

	\begin{figure*}[tb]
		\centering
		\includegraphics[width=171mm]{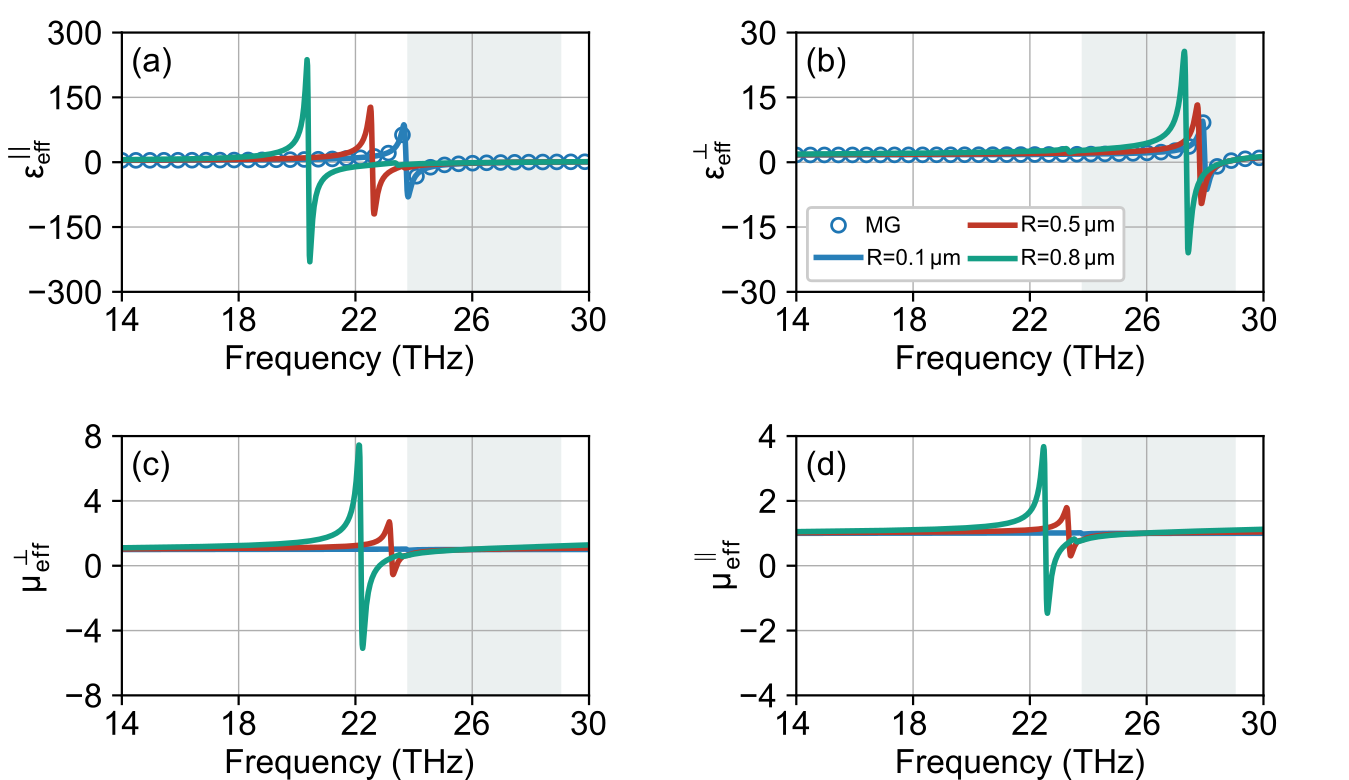}
		
		\caption{\label{fig:SiCeffepsilon} Parallel ((a),(d)) and perpendicular ((b), (c)) components of the relative effective permittivity (upper row) and  permeability (lower row) for SiC cylinders with filling ratio 30\% in air, for different radii (mentioned in the legends)  using the CPA (lines) and the Maxwell-Garnett approximation (circles). The shaded areas correspond to the frequency region where the dielectric function of SiC  is negative.   }
	\end{figure*}
	
	\begin{figure*}
		\centering
		\includegraphics{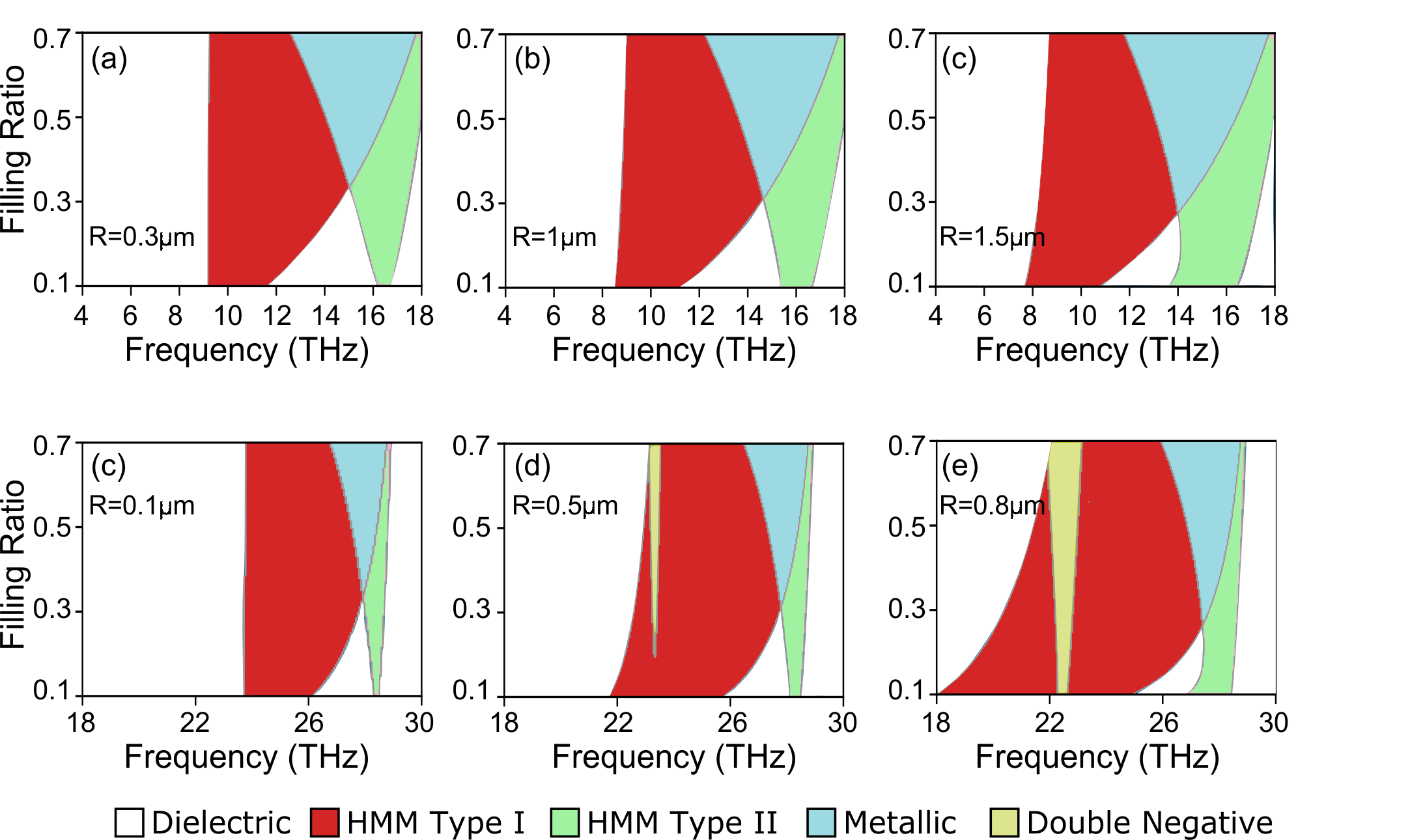}
		
		\caption{\label{fig:LiFoptical Phase} Optical phase diagrams for LiF (top row) and SiC (bottom row) cylinders in air for different radii, $R$, of the cylinders. The color marks the different attainable metamaterial-related responses of the systems:  Dielectric (White color): Both $\varepsilon^{\parallel}_\eff>0$, $\varepsilon^{\perp}_\eff>0$, $\mu^{\parallel}_\eff>0$, $\mu^{\perp}_\eff>0$, HMM Type I (red): $\varepsilon^{\parallel}_\eff<0$, $\varepsilon^{\perp}_\eff>0$, HMM Type II (green): $\varepsilon^{\parallel}_\eff>0$, $\varepsilon^{\perp}_\eff<0$, Metallic (blue): $\varepsilon^{\parallel}_\eff<0$, $\varepsilon^{\perp}_\eff<0$,     DNG (yellow): $\varepsilon^{\parallel}_\eff<0$,$\mu^{\perp}_\eff<0$.  }
	\end{figure*}
	We can now turn our attention to the calculation of effective medium material parameters $\varepsilon_{\eff}$ and $\mu_{\eff}$ for systems comprised of polaritonic cylinders in a host.  
	In Fig.~\ref{fig:LiFeffepsilon} we plot all the components of the effective permittivity and permeability tensors for LiF cylinders in air for the same set of radii discussed in  Section~\ref{sec:singlesc}, for single scattering, i.e. $0.3, 1$ and $1.5 \mum$, and LiF filling ratio $30\%$.
	We also plot the effective permittivities in the quasistatic limit ($k_h R\ll 1$ and $k_c R\ll 1$) using the Maxwell-Garnett approximation (see Eq.~(\ref{eq:MGappr}))~\cite{MGarnett1904RS}. 
	Since $\mu_h=\mu_c=1$ the effective magnetic permeabilities in the quasistatic limit are both equal to unity. 
	As one can see in Fig.~\ref{fig:LiFeffepsilon}, the effective permittivities and permeabilities exhibit Lorentzian-type resonances at frequencies close to their associated mode eigenfrequencies of a single cylinder, shown in Fig.~\ref{fig:LiFresonances}.  In a similar fashion to the single-cylinder eigenmodes, the resonances in the effective parameters move to lower frequencies and the maximum values of of $\varepsilon_{\eff}$ and $\mu_{\eff}$ increase for larger radii. 
	
	In particular, the so called electric modes, TM$_0$, TE$_1$,  lead to effective permittivity resonances, while the magnetic modes, TM$_1$, TE$_0$, to effective permeability resonances. Since the magnetic modes do not appear/resonate in the quasistatic regime (i.e. for small cylinder radii) the permeability resonances are not present in that regime, in agreement also with the MG formulation. In fact the accurate description and reproduction of magnetic effects in non-magnetic composites is one of the great merits of CPA regarding metamaterial effects and capabilities.
	
	Regarding the effective permittivity of Fig.~\ref{fig:LiFeffepsilon}, while for small cylinder radius the CPA results coincide with the Maxwell-Garnett results, as we increase the cylinder radius, exciting more resonances and thus more rich electromagnetic response, the Maxwell-Garnett is not able to describe the response of the inhomogenous medium and thus to reproduce the achievable metamaterial properties. A demonstration of this inability and the accuracy and success of our CPA approach is given in Fig.~\ref{fig:comparison}, where we compare the Maxwell-Garnett and the CPA results with full wave simulations for a polaritonic system that has been  realized also experimentally~\cite{Coronado2012OE}; that is a system of LiF cylinders (of radius $1.3\mu$m and filling ratio $6.95\%$) in a KCl host.
	\begin{comment}
	In particular, in  Fig.~\ref{fig:comparison} we compare the reflection coefficient for an LiF-KCl system of 7 unit cells along propagation direction (obtained through the commercial
	finite element method electromagnetic solver Comsol Multiphysics) with an homogeneous slab of the same thickness and parameters the effective parameters given by MG (green lines) and our CPA approach, for TM polarization (obtained through the Transfer Matrix method). 
	\end{comment}
	The results of Fig.~\ref{fig:SiCeffepsilon}, as well as analogous results for systems with smaller or larger cylinder radii, clearly demonstrate that 
	CPA can describe with satisfactory accuracy the electromagnetic response of  structures with larger-size cylinders (i.e. of $k_c R \approx 1)$, polaritonic or high-index-dielectric. (Note that the slight discrepancy between CPA and full-wave simulation results at ~8 THz is due to the fact that in this region $k_\eff^\TM R_2 \approx 3$, which is beyond the regime of validity of CPA.) 
	
	Coming back to our model systems, in Fig.~\ref{fig:SiCeffepsilon} we plot the components of the relative effective permittivity and permeability tensors for SiC cylinders in air for radii $0.1$, $0.5$ and $0.8 \mum$ (the same ones discussed in connection with Fig.~\ref{fig:SiCext}). The  filling fraction also here is chosen to be equal to $30\%$. 
	As in the case of LiF in air, we observe also here resonant permittivity and permeability, closely connected with single cylinder resonances, as discussed in the case of Fig.~\ref{fig:LiFeffepsilon}. A significant difference here is the stronger magnetic response leading to even negative permeability values; this is a result of the higher permittivity values of SiC compared to LiF (compare the permittivity values of Fig.~\ref{fig:LiFSiC}(a) and \ref{fig:LiFSiC}(b)),  and thus of the stronger displacement current.
	
	A closer examination of Figs.~\ref{fig:LiFeffepsilon} and~\ref{fig:SiCeffepsilon} indicates that there is a variety of interesting and useful metamaterial properties achievable by our polaritonic rod systems. These include (a) engineerable permittivity response comprising of both high positive values, negative values, and near-zero values; (b) engineerable permeability, including negative permeability values; (c) double-negative response, i.e. permittivity and permeability both negative, resulting to negative refractive index response, (d) hyperbolic response. Below we comment in more detail on the above properties and response, generalizing to any polaritonic-rod-based composite. Moreover in Appendix~\ref{app:SiCeffIm} we show also the imaginary parts of the effective permittivity and permeability components shown in Figs. 6 and 8, essential for an evaluation on the functionability of the corresponding systems/metamaterials.
	
	A. {\em Engineerable permittivity response}:
	Although in the bulk polaritonic materials we already have a rich permittivity response, including both positive, negative and near-zero values, structuring the polaritonic material in the form of cylinders we have the potential to engineer the permittivity values, reaching negative values even below the resonance of the corresponding bulk material (compare, e.g., Fig. 1(a) and Fig. 6(a) or Fig. 1(b) and Fig. 8(a)), reaching {\em desired} negative (or positive) values different than those of the bulk material as, e.g., to, adjust the impedance of the system with that of its surrounding medium, combining properly the real and imaginary parts of the effective $\varepsilon$ as, e.g., to reduce losses in the region of operation, moving the epsilon-near-zero response in the desired frequency range, etc. The effective permittivity values can be engineered by changing either the cylinders radii or the cylinders filling ratio.
	
	B. {\em Engineerable permeability response}:
	As Fig.~\ref{fig:SiCeffepsilon} shows, in properly designed systems of polaritonic rods, owing to the large permittivity values of the polaritonic materials, we have the ability to achieve resonant permeability associated with negative values for both TE and TM polarization if the underlying single-cylinder resonance is strong enough. The negative permeability response is favored by polaritonic materials of high $\varepsilon$ (compare the LiF with the SiC case), by cylinders of larger radii (as $k_c R \approx 1$) and by large cylinder filling ratio. As in the permittivity case, the effective permeability values can be engineered by changing either the cylinders radii or the cylinders filling ratio.
	
	C. {\em Double negative response}: 
	Regarding the double negative response resulting to negative refractive index, in the SiC system shown in Fig.~\ref{fig:SiCeffepsilon} we see that such a response is achievable (for TM polarization) for both $R=0.5\mu \text{m}$ and $R=0.8\mum$. (For $R=0.8\mum$ $\varepsilon^{\parallel}_\eff$ and $\mu^{\perp}_\eff$ are both negative between 22.2THz and 22.8THz). 
	Adjusting the cylinder radii or the filling ratio, one can engineer this response, engineering thus the effective impedance of the system and the effective refractive index. Having the potential to engineer separately refractive index and impedance offers a valuable tool for wave propagation  manipulation, as it allows perfect coupling to the surrounding medium or perfect transmission combined with desired phase propagation features.
	
	D. {\em Hyperbolic  response}:
	As can be seen in Figs. 6 and 8, for both the LiF and SiC systems studied here the condition for hyperbolic medium $\varepsilon^{\parallel}_{\eff}\cdot\varepsilon^{\perp}_{\eff}<0$ (for positive $\mu_{\eff}$) can be easily  achieved in two different frequency regions  even in the quasistatic limit. In the first region, around  $\omega_T$ and near the TM$_0$ resonance frequency, the out-of-plane components  $\varepsilon^{\parallel}_{\eff}$ are negative for all radii (at least for $R>0.1\mum$)  while the in-plane components $\varepsilon^{\perp}_{\eff}$ are positive; 
	the medium in this case is called \textit{Hyperbolic Medium Type I (HMM I)}. 
	The second frequency region where hyperbolic dispersion is feasible is near the TE$_1$ resonance frequency. Here the in-plane components $\varepsilon^{\perp}_{\eff}$  becomes negative while the other two components are positive; in this case the effective medium is called \textit{Hyperbolic Medium Type II (HMM II)}. 
	
	\subsubsection{Optical Phase Diagrams - Filling Ratio Influence}
	\label{subsc:opticalphase} 
	To illustrate further and more clearly the different attainable properties and capabilities of systems of polaritonic rods, we investigate for our two systems the frequency regions where the above mentioned interesting MM responses occur as we change the rods filling ratio. In Fig.~\ref{fig:LiFoptical Phase} we plot for both systems (i.e. LiF and SiC) the optical phase diagrams, showing the different interesting optical response regions as a function of filling ratio and frequency, for various radii. There the different regions are marked with different colors: With red the HMM I region, with green the HMM II, with yellow the DNG region (achievable for TM polarization) and with blue the fully metallic region ($\varepsilon_\eff^\perp<0$ and $\varepsilon_\eff^\parallel<0$).
	We should note here that for SiC there are regions with hyperbolic response in both electric permittivity $\varepsilon_\eff$ (where $\mu_{\eff}>0$) and magnetic permeability $\mu_\eff$ (where $\varepsilon_\eff>0$).
	For simplicity we don't separate these areas in our plots. 
	As one can see in Fig.~\ref{fig:LiFoptical Phase}, there is a pattern on the achievable response: high-$\varepsilon$ region at low frequencies is followed by a HMM I region starting near the optical phonon frequency $\omega_{T}$ and extending into the reststrahlen  band ($\omega_T<\omega<\omega_L$).
	Moreover, for high frequencies ($\omega>\omega_T$) and high filling ratios there is a region (green areas in Fig.~\ref{fig:LiFoptical Phase}) where the material exhibits a purely metallic response (both $\varepsilon^\parallel_\eff<0$ and $\varepsilon^\perp_\eff<0$). 
	In addition, the boundaries between the different optical phases correspond to frequencies where epsilon-near-zero is achievable since one or more of the components of the effective permittivity changes sign. 
	Finally, the DNG region (achievable for TM polarization) is always inside the HMM I region and requires polaritonic materials of high permittivity values (i.e. of strong phonon-polariton resonance) and not extremely subwavelength in size cylinders; moreover it is favored from larger cylinder filling ratios. 
	%For instance, for $R=0.8\mum$, the system exhibits DNG response with with frequency range  up to $1.2\THz$. 

	% CONCLUSIONS
	\section{Conclusions}
	\label{sec:concl}
	Prompted by the constantly growing interest on polaritonic and dielectric metamaterials, we presented here a detailed study  of the electromagnetic response of metamaterial systems formed by polaritonic rods in a dielectric host, in the THz region of the electromagnetic spectrum. Employing as model systems systems of LiF and SiC rods, we initially studied the response of single rod and we calculated the extinction efficiency for different radii of the rod, in order to identify the nature and behavior of the major resonances for each polarization. Subsequently, using the single-rod scattering formulation and data and employing the Coherent Phase Approximation effective medium approach, which can accurately describe an inhomogeneous medium even beyond the quasistatic regime, we obtained closed formulas for the effective parameters of systems made of polaritonic rods in a host and we applied them in  the cases of LiF and SiC rods. 
	We found that by proper selection of the radius and the filling ratio of the rods one can achieve a variety of interesting and useful metamaterial properties in polaritonic rod systems. These properties include engineerable permittivity (having high positive, negative and near-zero values), engineerable permeability (of both positive and negative values), hyperbolic response,  double negative response and others.
	The possibility to achieve this rich variety of physical properties in the THz region, which is of high technological interest, combined with the ease of fabrication of many of those systems, makes polaritonic rod metamaterials ideal candidates for any device aiming THz wave propagation and scattering control.

	% ACKNOWLEDGEMENTS
	\section{Acknowledgements}
	We acknowledge financial support by  the European Union’s Horizon
	2020 FETOPEN programme under projects VISORSURF (grant agreement No.
	736876) and NANOPOLY (grant agreement No. 829061), and by the
	General Secretariat for Research and Technology, and the
	H.F.R.I. Ph.D. Fellowship Grant (Grant Agreement No. 4894)
	in the context of the action “1st Proclamation of Scholarships
	from ELIDEK for Ph.D. Candidates.”
	
	% 
	% APPENDICES 
	%
	\appendix
	\section{Effective Medium Derivation}
	\label{app:affective-Derivation} 
	In this appendix we  give a brief derivation of the relations for the effective medium. 
	We consider a coated cylinder (along ${\hat z}$ direction) with core radius $R_1$ and  shell thickness $R_2-R_1$ embedded in an infinite medium with material parameters $\varepsilon_{\eff},\mu_{\eff}$. 
	The system is shown in Fig.~\ref{fig:coated}(c).
	The core cylinder is made of a material with material parameters $\varepsilon_{c},\mu_{c}$ and the shell of a material with material parameters $\varepsilon_{h},\mu_{h}$.
	
	Depending on the incident wave polarization the fields (in G-CGS system of units) in each region can be expanded on the appropriate cylindrical harmonics $\mathbf{N}_{e m k}=k Z_m(k\rho)\cos(m\varphi)\hat{z}$ and $\mathbf{M}_{e m k} = -\frac{m}{\rho}Z_m(k\rho)\sin(m\varphi)\hat{\rho}-kZ'_m(k\rho)\cos(m\varphi)\hat{\varphi}$, with $k$ the wavenumber,  $Z_m(k\rho)=H_m(k\rho)$ for outward-going waves and $Z_m(k\rho)=J_m(k\rho)$ for inward-going waves ~\cite{Stratton2015Book}. For TE polarization (electric field perpendicular to the cylinder axis) the fields outside the coated cylinder are a sum of the incident (inward) and scattered (outward) fields, and can be expressed as   
	\begin{align}
	\label{eq:E1}
	\mathbf{E}_{\text{out}} &= i\sum_{m=0}^{\infty} A_{m k_\eff} \left[ D_m^{\TE} \mathbf{M}_{emk_\eff}^{(\text{outward})} +\mathbf{M}_{e m k_\eff}^{(\text{inward})} \right] \\
	\label{eq:Hz1}
	\mathbf{H}_{\text{out}} &= \frac{c k_\eff}{\omega \mu_\eff}\sum_{m=0}^{\infty} A_{mk_\eff} \left[ D_m^{\TE} \mathbf{N}_{emk_\eff}^{(\text{outward})} +\mathbf{N}_{e m k_\eff}^{(\text{inward})} \right] 
	\end{align}
	\begin{comment}
	\label{eq:E2}
	\mathbf{E}_{\text{host}} &=& i\sum_{m=0}^{\infty} A_{m k_\eff} \left[ C_m^{\TE \text{(out)}} \mathbf{M}_{emk_\eff}^{(\text{out})} +C_m^{\TE \text{(in)}}\mathbf{M}_{e m k_\eff}^{(\text{in})} \right] \\
	\mathbf{H}_{\text{host}} &=& \frac{c k_h}{\omega \mu_h}\sum_{m=0}^{\infty} A_{m k_h} \left[ C_m^{(out)} \mathbf{N}_{em k_h}^{\TE (\text{out})} + C_m^{\TE \text{(in)}} \mathbf{N}_{em k_h}^{(\text{in})} \right] \\
	\label{eq:E3}
	\mathbf{E}_{\text{cyl}} &=& i\sum_{m=0}^{\infty} A_{m k_c}  B_m^{\TE} \mathbf{M}_{e\nu k_c}^{(\text{in})}\\
	\label{eq:Hz3}
	\mathbf{H}_{\text{cyl}} &=& \frac{c k_c}{\omega \mu_c}\sum_{m=0}^{\infty} A_{m k_c}  B_m^{\TE} \mathbf{N}_{e\nu k_c}^{(\text{in})}
	\end{comment}
	where $D^{\TE}_m$ are the scattering coefficients 
	%, $B^{\TE}_m$, $C^{\TE (\text{in})}_m$, and $C^{\TE (\text{out})}_m$ denote the appropriate expansion coefficients 
	and $A_{mk} = \frac{1}{k}\frac{2}{1+\delta_{m 0}} i^m$. 
	In an analogous way one can express the fields in all the regions of the scattering system, i.e. in the core cylinder and the coating.
	\begin{figure}[tb]
		\centering
		\includegraphics[width=86mm]{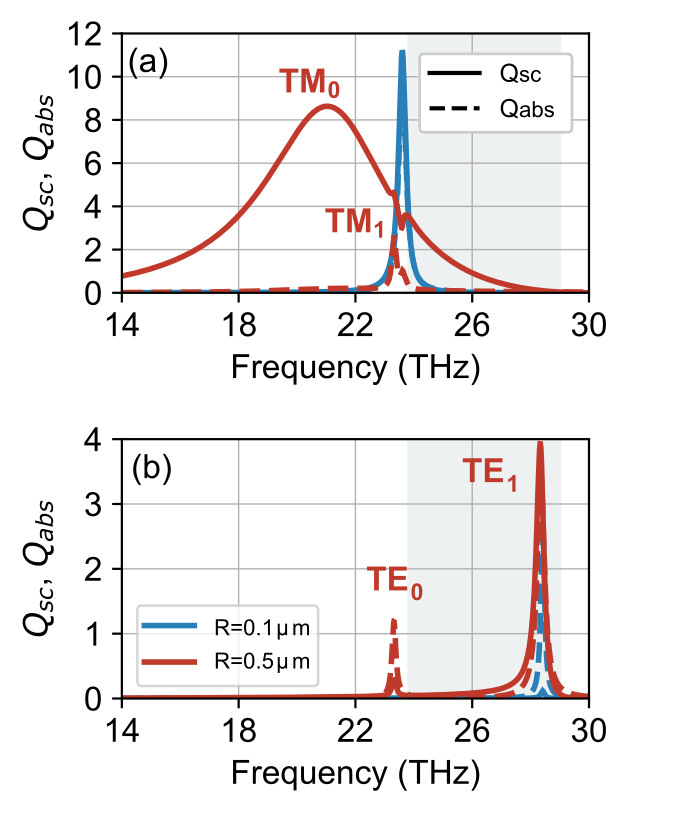}
		
		\caption{\label{fig:LiFabssc} Scattering (solid lines) and absorption (dashed lines) efficiencies of a LiF cylinder (of radius 0.3 and 1.5 $\mu$m) in air for (a) TM and (b) TE polarization. The shaded areas correspond to the frequency region where the dielectric function of LiF is negative. }
	\end{figure}
	\begin{figure}[tb]
		\centering
		\includegraphics[width=86mm]{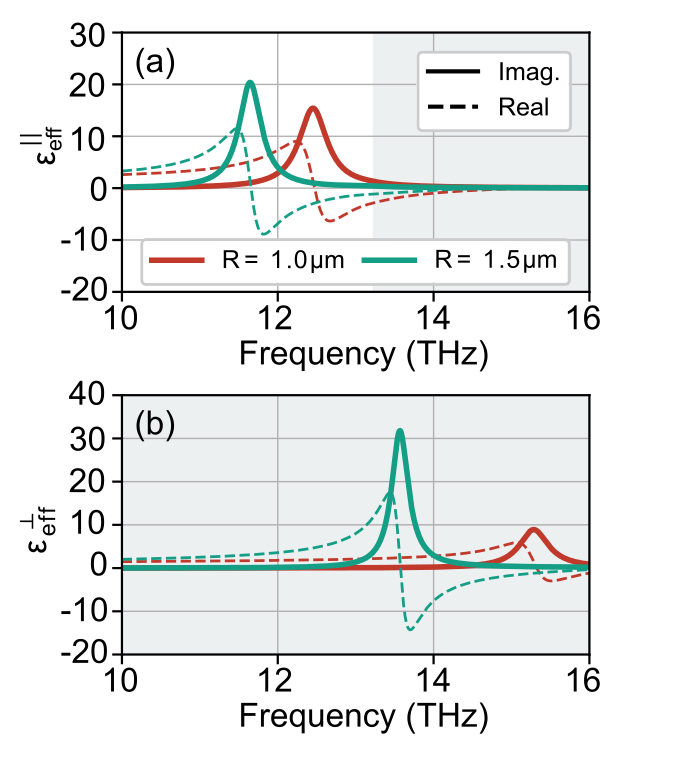}
		
		\caption{\label{fig:SiCabssc} Scattering (solid lines) and absorption (dashed lines) efficiencies of a SiC cylinder (of radius 0.1 and 0.5 $\mu$m) in air for (a) TM and (b) TE polarization. The shaded areas correspond to the frequency region where the dielectric function of SiC is negative. }
	\end{figure}
	\begin{figure}[tb]
		\centering
		\includegraphics[width=86mm]{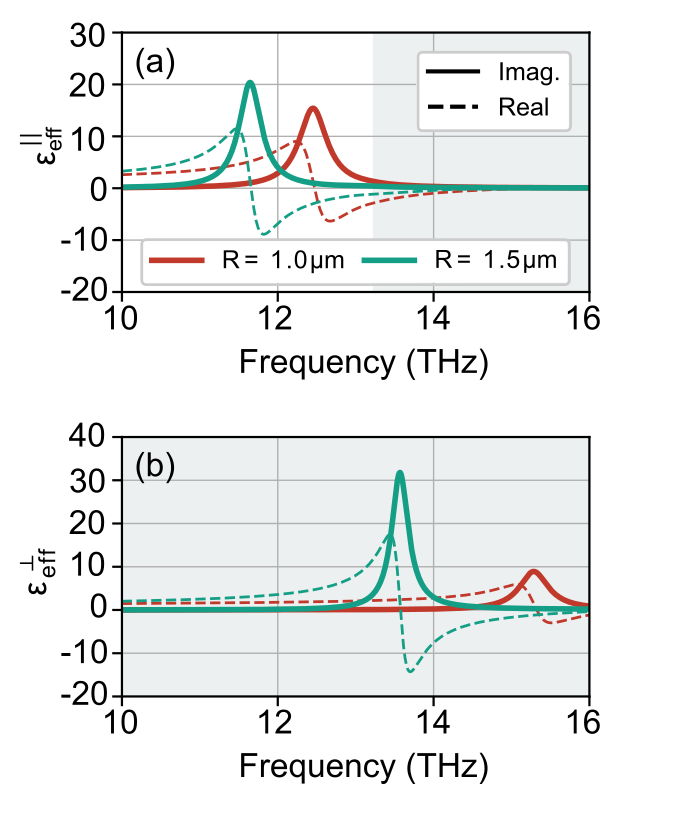}
		
		\caption{\label{fig:LiFeffIm} Imaginary (solid lines) and real (dashed lines) parts of the parallel (a) and the perpendicular (b) components of the relative effective permittivity for LiF cylinders with filling ratio 30\% in air, for different radii, $R$ (mentioned in the legends),  using the CPA. The shaded areas correspond to the frequency region where the dielectric function of LiF  is negative.  }
	\end{figure}
	The scattering coefficients and all the coefficients appearing in the expansion of the fields in cylindrical harmonics can be obtained by imposing the appropriate boundary conditions at the different system interfaces.
	Applying those conditions one can find that 
	the scattering coefficient $D_m^\TE$ take the form
	\begin{equation}
	\label{eqB:anTE}
	D_m^{\text{TE}} = \frac{n_h\mu_{\eff} \mathcal{T}_m J'_m (k_{\eff}R_2)-n_{\eff} \mu_h \mathcal{K}_m J_m (k_{\eff}R_2)}{n_{\eff}\mu_h \mathcal{K}_m H_m (k_{\eff}R_2)-n_h \mu_{\eff} \mathcal{T}_m H'_m (k_{\eff}R_2)}
	\end{equation}
	where
	\begin{gather}
	\mathcal{K}_m = \Theta_m H'_m (k_h R_2) + \Phi_m J'_m(k_h R_2) \\
	\mathcal{T}_m = \Theta_m H_m (k_h R_2) + \Phi_m J_m (k_h R_2)
	\end{gather}
	with
	\begin{gather}
	\Theta_m = n_c \mu_h J'_m(k_h R_1) J_m (k_c R_1) - n_h \mu_c J_m(k_h R_1)J'_m (k_cR_1) \\
	\Phi_m = n_h \mu_c H_m(k_h R_1) J'_m (k_cR_1) - n_c \mu_h H'_m(k_h R_1)J_m (k_cR_1) 
	\end{gather}
	and $\Theta_m/\Phi_m = a^{\text{TE}}_m(R_1;c,h)$ are the scattering coefficients of a single cylinder of radius $R_1$ with material parameters $\varepsilon_c$ and $\mu_c$ embedded in a host material of parameters $\varepsilon_h$ and $\mu_h$, i.e. a cylinder of the original system to be homogenized.
	
	For TM polarization the fields outside the coated cylinder can be expressed as
	\begin{align}
	\label{eq:E2tm}
	\mathbf{E}_{\text{out}} &= \sum_{m=0}^{\infty} A_{m k_\eff} \left[ D_m^{\TE} \mathbf{N}_{emk_\eff}^{(\text{outward})} +\mathbf{N}_{e m k_\eff}^{(\text{inward})} \right] \\
	\label{eq:Hz2tm}
	\mathbf{H}_{\text{out}} &= i\frac{c k_\eff}{\omega \mu_\eff}\sum_{m=0}^{\infty} A_{mk_\eff} \left[ D_m^{\TM} \mathbf{M}_{emk_\eff}^{(\text{outward})} +\mathbf{M}_{e m k_\eff}^{(\text{inward})} \right] 
	\end{align}
	with the $m$th-order scattering coefficient given by  
	\begin{equation}
	\label{eqB:bnTM}
	D^{\TM}_m =  \frac{n_h\mu_{\eff} \mathcal{U}_m J_m (k_{\eff}R_2)-n_{\eff} \mu_h \mathcal{Y}_m J'_m (k_{eff}R_2)}{n_{\eff}\mu_h \mathcal{Y}_m H'_m (k_{\eff}R_2)-n_h \mu_{\eff} \mathcal{U}_m H_m (k_{\eff}R_2)}
	\end{equation}
	where
	\begin{align}
	\mathcal{Y}_m = \Pi_m H_m (k_h R_2) + \Lambda_m J_m (k_h R_2) \\
	\mathcal{U}_m = \Pi_m H'_m (k_h R_2) + \Lambda_m J'_m (k_h R_2)
	\end{align}
	and
	\begin{align}
	\Lambda_m = n_h \mu_c H'_m (k_h R_1) J_m (k_c R_1) - n_c \mu_h H_m (k_h R_1)J'_m (k_cR_1) \\
	\Pi_m = n_c \mu_h J_m(k_h R_1) J'_m (k_c R_1) - n_h \mu_c J'_m(k_h R_1)J_m (k_cR_1) 
	\end{align}
	where $\Pi_m/\Lambda_m = a^{\text{TM}}_m(R_1;c,h)$.
	\begin{figure*}[tb]
		\centering
		\includegraphics[width=171mm]{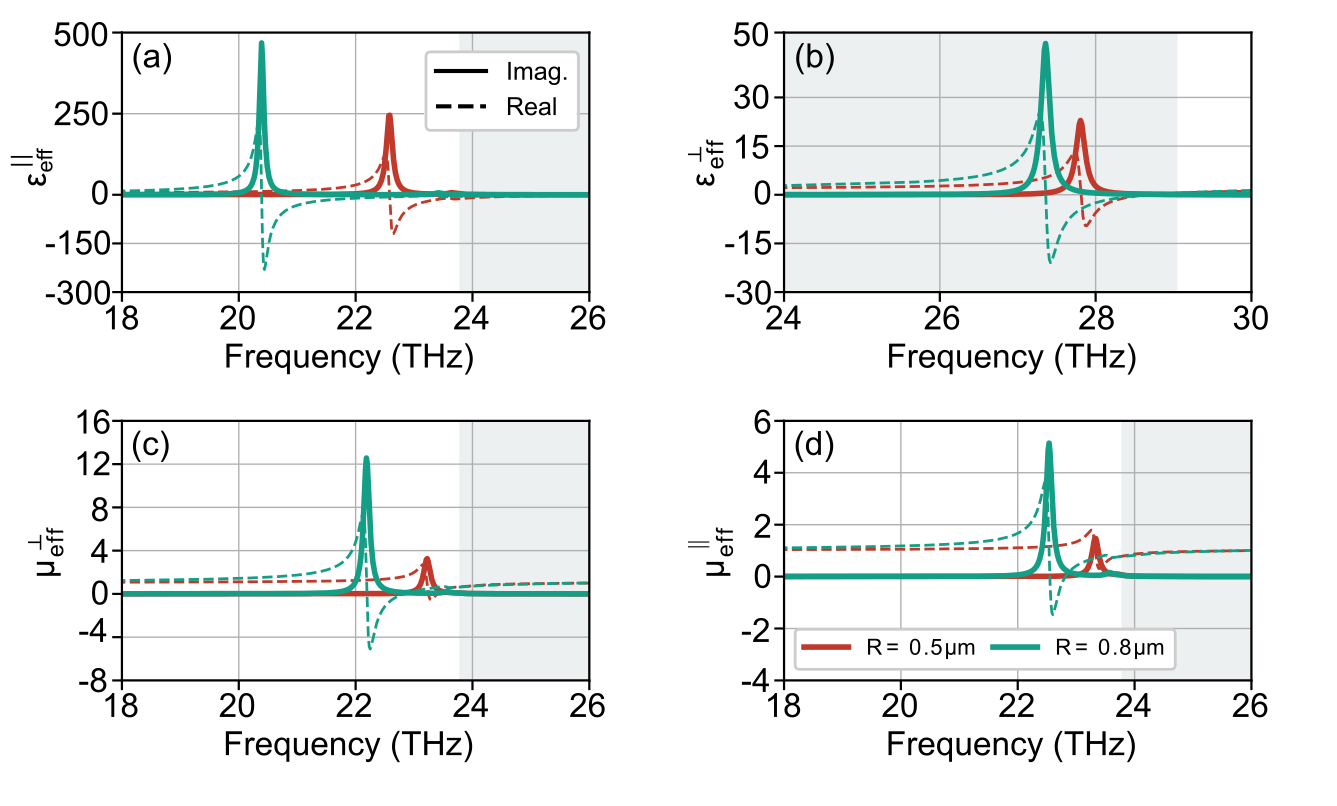}
		
		\caption{\label{fig:SiCeffIm} Imaginary (solid lines) and real (dashed lines) parts of parallel ((a),(d)) and perpendicular ((b), (c)) components of the relative effective permittivity (upper row) and  permeability (lower row) for SiC cylinders with filling ratio 30\% in air, for different radii, $R$ (mentioned in the legends)  using the CPA. The shaded areas correspond to the frequency region where the dielectric function of SiC  is negative.   }
	\end{figure*}
	Following the CPA main concept, for the medium hosting the coated inclusion to be the valid effective medium (i.e. the medium approximating the original system of cylinders of $\varepsilon_c$ and $\mu_c$ in the host of $\varepsilon_h$ and $\mu_h$), we must require the scattering cross section from the coated inclusion to be identically zero. Hence, all the scattering coefficients must be set equal to zero.  That is, 
	\begin{equation}
	D_m^{\mathtt{P}} = 0
	\end{equation}
	where $\mathtt{P}=\lbrace\TM, \TE\rbrace$. This equation reduces to the much simpler one, that is 
	\begin{equation} 
	\label{eq:CPA_TE}
	a_m^{\mathtt{P}}(R_1;c,h) =  a_m^{\mathtt{P}}(R_2;\eff,h)
	\end{equation}
	In Eq.~(\ref{eq:CPA_TE}) we have only the coefficients of simple (non-coated) cylinders (given by Eqs. ~(\ref{eq:Hcoeff}) and (\ref{eq:Ecoeff})), since the scattering coefficient  $a_m^{\mathtt{P}}(R;A,B)$ denotes the $m$-th order  coefficient for a cylinder of radius $R$ and material parameters $\varepsilon_A$, $\mu_A$ embedded in a medium with $\varepsilon_B$, $\mu_B$.
	
	If we consider only the $m=0$ and $m=1$ terms in Eq.~(\ref{eq:CPA_TE}), which are the dominant terms in the long-wavelength limit, we can find explicit relations for all components of the permittivity and permeability tensors in the region $k_{\eff}R_2 < 1$. To do so we replace the Bessel functions with argument $k_{\eff}R_2$ (of order 0 and 1 and their derivatives) by their limiting expressions for small argument, employing the  series expansions
	\begin{align}
	J_0(x) &\approx 1-\frac{x^2}{4} \label{eq:besselLim0} \\
	-J_1(x) = J'_0(x)  &\approx  -\frac{x}{2}+\frac{x^2}{16} \label{eq:besselLim1} \\
	J'_1(x) &\approx  \frac{1}{2} - \frac{3}{16}x^2 \label{eq:besselDLim1} \\
	H_0(x) &\approx \frac{2i}{\pi} \left[\ln(x/2)+\gamma\right]+1 \label{eq:hankelLim0}\\ 
	-H_1(x) = H'_0(x)  &\approx \frac{2i}{\pi x} - \frac{x}{2}+i\alpha x \label{eq:hankelLim1} \\
	H'_1(x) &\approx  \frac{2i}{\pi x^2} + \frac{1}{2} +\frac{i}{\pi} - i\alpha \label{eq:hankelDLim1} 
	\end{align} 
	where $\gamma= 0.577215$ is the Euler-Mascheroni constant and $\alpha=-\frac{1}{\pi}\left[\ln(x/2)+\gamma-\frac{1}{2}\right]$.

	Employing Eqs.~(\ref{eq:CPA_TE}) and (\ref{eq:besselLim0})-(\ref{eq:hankelDLim1}) (keeping in most of the cases only their lowest order term) we result to the effective medium formulas (\ref{eq:cpa1})-(\ref{eq:cpa4}) of the main text.
	In particular,
	\begin{eqnarray}
	a_0^{\TM}(R_1;c,h) =  a_0^{\TM}(R_2;\eff,h) & \rightarrow & \varepsilon_\eff^\parallel \\
	a_1^{\TM}(R_1;c,h) =  a_1^{\TM}(R_2;\eff,h) & \rightarrow & \mu_\eff^\perp \\
	a_0^{\TE}(R_1;c,h) =  a_0^{\TE}(R_2;\eff,h) & \rightarrow & \mu_\eff^\parallel \\
	a_1^{\TE}(R_1;c,h) =  a_1^{\TE}(R_2;\eff,h) & \rightarrow & \varepsilon_\eff^\perp 
	\end{eqnarray}

	\section{Scattering/absorption}
	\label{app:SiCscabs}
	To illustrate more clearly the dependence of the single-cylinder absorption and scattering efficiencies  on the cylinder radius, which was discussed in connection with Fig. 5, we present here  the scattering and absorption efficiencies for different indicative radii. Fig.~\ref{fig:LiFabssc} shows the scattering and absorption efficiencies for a LiF cylinder in air, while Fig.~\ref{fig:SiCabssc} shows corresponding results for a SiC cylinder. As can be seen in Figs.~\ref{fig:LiFabssc} and Fig.~\ref{fig:SiCabssc}, the results support  the discussion of Section III.A regarding the tendencies of the scattering and absorption efficiencies as the cylinder radius increases. 
	
	\section{Effective parameters}
	\label{app:SiCeffIm} 
	In Figs.~\ref{fig:LiFeffIm} and \ref{fig:SiCeffIm} we plot the imaginary part of the effective permittivity and permeability components for the systems of LiF and SiC cylinders discussed in Section~\ref{subsc:effective}. We also plot there the real part of those components, copied from Figs.~\ref{fig:LiFeffepsilon} and \ref{fig:SiCeffepsilon},  for an easy comparison and assessment of the functionality of the composites. As can be  seen from Figs.~\ref{fig:LiFeffIm} and \ref{fig:SiCeffIm}, apart of a very narrow frequency region around the resonance frequencies of the components where the losses are  significant, in all other frequency regions the losses are quite negligible. This shows that, unlike many plasmonic systems, in polaritonic rod systems resistive losses are not a major problem hindering their applicability.
	
	% REFERENCES
	%\bibliography{polaritonic} 
	
	%\bibliographystyle{phys}

\end{document}